\documentclass[twocolumn,pra,aps,superscriptaddress]{revtex4-1}
\usepackage{amssymb,amsmath,amsfonts}

\usepackage[latin9]{inputenc}
\usepackage{color}
\usepackage{amstext}
\usepackage{graphicx}
\usepackage[unicode=true,pdfusetitle,bookmarks=false,breaklinks=false,pdfborder={0 0 1},backref=false,colorlinks=false]{hyperref}
\hypersetup{bookmarksnumbered=false,bookmarksopen=false}

\usepackage{hyperref}
\hypersetup{colorlinks,citecolor=NavyBlue,filecolor=black,linkcolor=black,urlcolor=black}
\usepackage[dvipsnames]{xcolor}

\makeatletter

\newcommand\adag{a^\dagger}
\newcommand\adaga{a^\dagger a}

\newcommand\ket[1]{\left|#1\right\rangle}
\newcommand\bra[1]{\left\langle #1 \right|}

\usepackage{times}

\makeatother

\begin{document}
\title{Quantum simulation of multiphoton and nonlinear dissipative spin-boson models}

\author{R. Puebla}\email{r.puebla@qub.ac.uk}
\affiliation{Centre for Theoretical Atomic, Molecular, and Optical Physics, School of Mathematics and Physics, Queen's University, Belfast BT7 1NN, United Kingdom}

\author{J. Casanova}
\affiliation{Department of Physical Chemistry, University of the Basque Country UPV/EHU, Apartado 644, E-48080 Bilbao, Spain}
\affiliation{IKERBASQUE, Basque Foundation for Science, Maria Diaz de Haro 3, 48013 Bilbao, Spain}
\author{O. Houhou}
\affiliation{Centre for Theoretical Atomic, Molecular, and Optical Physics, School of Mathematics and Physics, Queen's University, Belfast BT7 1NN, United Kingdom}
\affiliation{Laboratory of Physics of Experimental Techniques and Applications, University of M\'ed\'ea, M\'ed\'ea 26000, Algeria}

\author{E. Solano}
\affiliation{Department of Physical Chemistry, University of the Basque Country UPV/EHU, Apartado 644, E-48080 Bilbao, Spain}
\affiliation{IKERBASQUE, Basque Foundation for Science, Maria Diaz de Haro 3, 48013 Bilbao, Spain}
\affiliation{Department of Physics, Shanghai University, 200444 Shanghai, China}

\author{M. Paternostro}
\affiliation{Centre for Theoretical Atomic, Molecular, and Optical Physics, School of Mathematics and Physics, Queen's University, Belfast BT7 1NN, United Kingdom}

\begin{abstract}
 We present a framework for the realization of dissipative evolutions of spin-boson models, including multiphoton exchange dynamics, as well as nonlinear transition rates. Our approach is based on the implementation of a generalized version of a dissipative linear quantum Rabi model.  The latter comprises a linearly coupled spin-boson term, spin rotations,  and standard dissipators. We provide numerical simulations of illustrative cases supporting the good performance of our method. Our work allows for the simulation of a large class of fundamentally different quantum models where the effect of distinct dissipative processes can be easily investigated.
\end{abstract}

\maketitle

\section{Introduction}\label{sec:intro}

The interaction between a bosonic mode and a two-level system is one of the most fundamental models in quantum physics and, consequently, is of significant relevance in several branches  of modern science such as quantum information~\cite{Nielsen} and light-matter interaction theory~\cite{Scully}. Here, the  Rabi model~\cite{Rabi:36,Rabi:37} and its simplified but quantized version known as the Jaynes-Cummings model (JCM)~\cite{Jaynes:63} play a prominent role. The quantum Rabi model (QRM) describes the coherent exchange of excitations between a spin and a bosonic mode, and despite its simplicity displays a rich variety of physical phenomena. As a matter of fact, during the last decade this model has attracted great interest from different research areas~\cite{Solano:11,Braak:16}. The diversity of physical phenomena encompassed in this model and its scientific relevance are embodied by the paradigmatic Rabi oscillations (or simply continuous revivals of quantum populations), its integrability~\cite{Braak:11}, the emergence of distinct behavior in the ultra-strong~\cite{Ciuti:05,Anappara:09,Forn:17,Kockum:18} and deep-strong coupling regimes~\cite{Casanova:10}, and by the existence of quantum phase transitions in a suitable parameter limit~\cite{Hwang:15,Puebla:16,Puebla:17,Liu:17}.

The QRM can be achieved in a variety of quantum platforms, being realizable in trapped ions~\cite{Pedernales:15,Lv:18}, circuit-QED~\cite{Ballester:12,Mezzacapo:14,Langford:17,Braumueller:17}, cold atoms~\cite{Schneeweiss:18}, spin-mechanical systems~\cite{Abdi:17} and integrated optics~\cite{Crespi:12}. However, the disparate ways in which a spin and a bosonic mode can interact goes certainly beyond the realm of the QRM. In this respect, one can find models possessing multiphoton exchange dynamics~\cite{Brune:87,Toor:92,Felicetti:15,Duan:16,Puebla:17pra} and/or nonlinear transition rates~\cite{MatosFilho:94,MatosFilho:96,Vogel:95}, which may unveil novel and interesting phenomena. Remarkably, while a QRM can typically be well experimentally realized, the implementation of other interaction mechanism such as those involving spin-boson nonlinear terms remains a challenging task.

From a different perspective, models comprising $n$-boson excitation-exchange processes with a spin degree of freedom have been theoretically studied mainly in their $n=2$ form, i.e. in the so-called two-photon QRM (2QRM)~\cite{Felicetti:15,Duan:16,Puebla:17pra}. The 2QRM is of particular interest for preparing non-classical states of light~\cite{Brune:87,Toor:92}, while  its solvability has been also studied~\cite{Travenec:12,Chen:12,Peng:13,Cui:17}. In addition, it is worth mentioning that systems comprising both one- and two-boson exchange terms can be as well of interest~\cite{Bertet:05,Felicetti:18,Ying:18}, even for  the simulation of relativistic effects~\cite{Pedernales:18}. Furthermore, these multiphoton models can be classified as either linear or nonlinear, that is, depending on whether their spin-boson coupling changes with the Fock occupation number~\footnote{A remark is due to clarify the terminology that will be used throughout this paper: although multiphoton spin-boson models such as the 2QRM do involve  \textit{nonlinear} interactions among bosonic and spin operators, we will refer to nonlinear interactions whenever the spin-boson coupling constant depends in a nonlinear manner on the Fock occupation number, regardless of the number of exchanged bosons. Such nonlinear models, originally proposed to give account for the quantum dynamics of a trapped-ion beyond the Lamb-Dicke regime~\cite{MatosFilho:94,MatosFilho:96,Vogel:95}, may have also applications in the simulation of Franck-Condon physics~\cite{Hu:11} or in dissipative preparation of Fock states~\cite{Cheng:18}}.

It is also worth mentioning that  the interaction of a single spin with a large  number (typically infinite) of harmonic oscillators~\cite{Weiss}, has served as a test-bed to scrutinize aspects of quantum dissipation due to the presence of an environment.  As a microscopic description of such dissipative effects is often very demanding (if not unfeasible)~\cite{Breuer}, it is customary to rely on a phenomenological description of the system-environment interaction based on a Lindbladian open-system framework~\cite{Breuer}. However, it is still possible to map a spin interacting with an infinite number of environmental harmonic oscillators into a generalized QRM, whose interaction with the environment is now mediated only through the bosonic mode~\cite{Thoss:01,Martinazzo:11,Iles:14,Iles:16,Nazir:18}. As recently proved in~\cite{Tamascelli:18}, under certain cases such method establishes an equivalence between non-Markovian dynamics of a spin immersed in a structured environment and a standard Markovian dynamics of a spin coherently coupled to a harmonic oscillator (see Refs.~\cite{Tamascelli:18} and~\cite{deVega:17} for the required conditions for an exact equivalence  and for a recent review on non-Markovian dynamics of open quantum systems, respectively). Moreover, it is worth stressing that Markovian dissipation may yield a dissipative phase transition in the QRM~\cite{Hwang:18} as well as a rich phenomenology when considering a large collection of spins~\cite{Morrison:08,Kirton:17,Kirton:18,Shammah:18,Soriente:18,Jauregui:18}. In short, studying and exploring the dynamics of a generalized QRM undergoing dissipation, even when it is of a Lindblad form, allows for the inspection of more complicated models and their interaction with  uncontrolled degrees of freedom that form the environment. 

Recently, it has been shown that a spin-boson model (for the case of having a single bosonic mode) comprising $n$-boson exchange terms can be realized only by having standard one-boson exchange terms plus spin rotations~\cite{Casanova:18}. In this manner, a model comprising only linear spin-boson couplings (a generalized QRM) allows to explore the fundamentally different physics of its multiphoton counterparts, such as that of a two- or three-photon QRM (2QRM) and (3QRM), respectively, without the need of implementing these experimentally challenging $n$-boson interaction terms that lead to multiphoton exchange. Our work goes beyond such previous results by {\em i)}~showing how standard dissipation translates into simulated multiphoton spin-boson models, leading to nontrivial dissipators; {\em ii)}~how nonlinear spin-boson models, as defined in~\cite{MatosFilho:94,MatosFilho:96,Vogel:95} i.e. those that emerge beyond the Lamb-Dicke regime, can be directly accessed with only linear interactions. We thus propose a novel strategy for the simulation of nonlinear models.  Moreover, as realizing a generalized QRM can be well attained in different platforms, we open new ways for the simulation of such nonlinear models in platforms that are relevant for quantum information processing --- such as microwave-driven ions~\cite{Woelk:17} or circuit-QED~\cite{Devoret:13,Gu:17} ---  but that are unsuited for the direct achievement of the desired nonlinearities.  

Our theoretical framework unveils a deep connection among such multiphoton and nonlinear models, which might have potential applications in quantum simulation and information processing, as well as in the inspection of the impact of dissipative processes in quantum optical processes.

The remainder of this article is organized as follows. In Sec.~\ref{sec:theory} we present the general theory, explaining the steps required to establish the approximate relation among the mentioned dissipative models. In Sec.~\ref{sec:res} we illustrate our theoretical apparatus by discussing specific cases in which a linear spin-boson model with typical dissipative processes realizes a nonlinear multi-boson model with transformed jump operators.  
We exemplify how such dissipative models can be realized to a very good approximation by simply using a generalized QRM, and validate our predictions  by performing detailed numerical simulations, as shown in Sec.~\ref{ss:num}.  Finally, in Sec.~\ref{sec:conc} we draw our conclusions and discuss possible additional directions of investigation.

\section{Theoretical framework}\label{sec:theory}
\begin{figure}
\centering
\includegraphics[width=0.9\linewidth,angle=00]{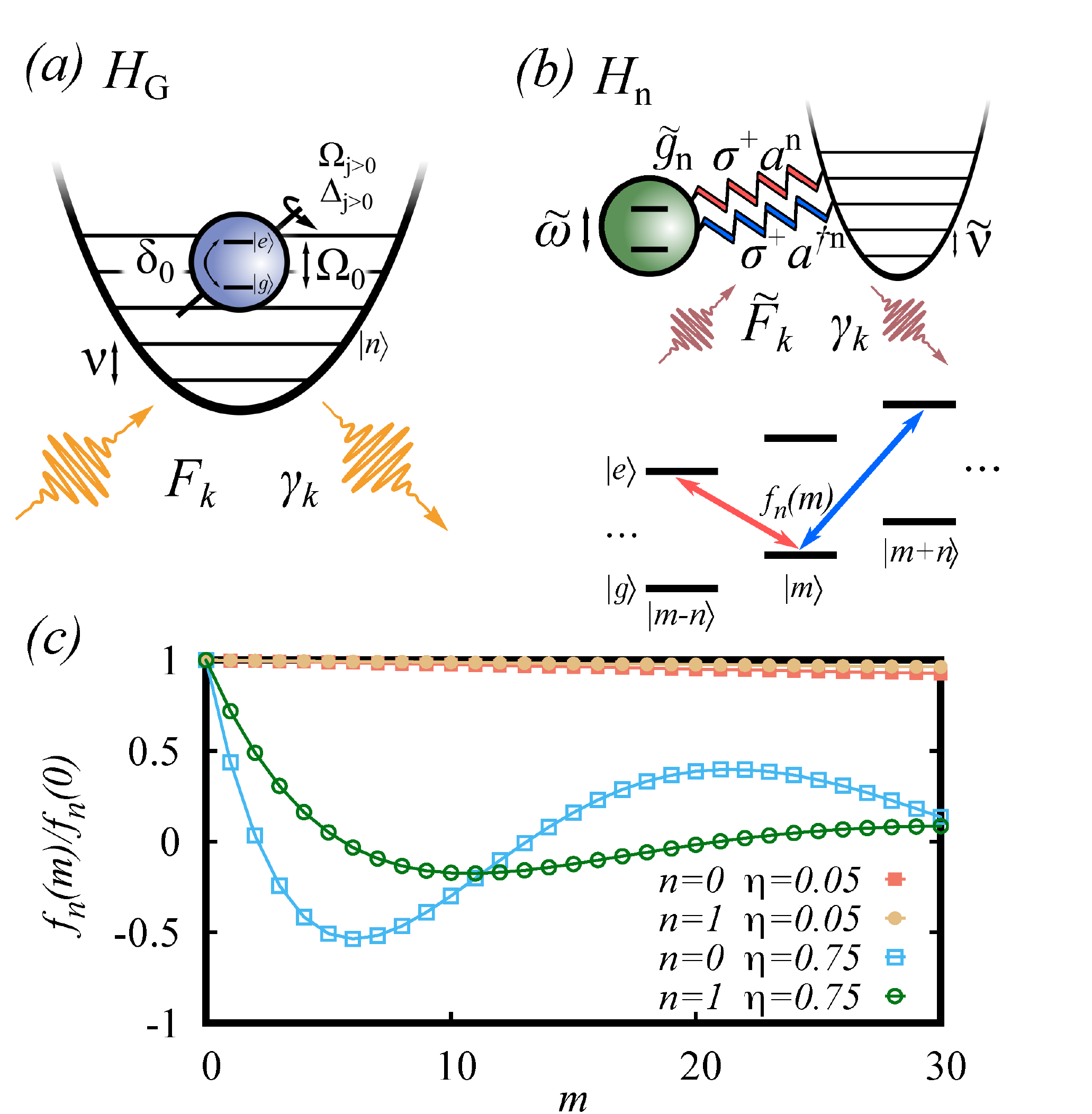}
\caption{\small{Sketch of the scheme for the simulation of nonlinear, multiphoton and dissipative spin-boson models. In panel {\it (a)} a linearly coupled spin-boson system (a generalized quantum Rabi model with spin-bias $\delta_0$, spin drivings with amplitude $\Omega_{j>0}$ and frequency $\Delta_{j>0}$) with Hamiltonian $H_{\rm G}$ [cf. Eq.~(\ref{eq:HG})] undergoes dissipation characterized by the rates $\gamma_k$ and the jump operators $F_k$. This allows us to simulate a nonlinear $n$-boson spin-boson model undergoing dissipation described by the $\tilde{F}_k$'s operators. In panel {\it (b)} the Hamiltonian is $H_{\rm n}$ [cf. Eq.~(\ref{eq:Hn})], which comprises interaction terms such as $\sigma^+a^n+{\rm H.c.}$  and/or $\sigma^+(\adag)^n+{\rm H.c.}$. In addition, the transition rates may strongly depend on the Fock-state label $m$ through a function $f_n(m)$ whose behavior is shown in panel {\it (c)} for different cases [cf. Sec.~\ref{sec:theory} and Eq.~(\ref{eq:fn})]. Note that we show $f_n(m)=\bra{m}f_n(\adaga)\ket{m}$, rescaled by $f_n(0)$, for various choices of $n$ and $\eta$. While $f_n(m)\approx f_n(0)$ for small values of $\eta$ and $m$, $f_n(m)$ significantly depends on $m$ for larger values of $\eta$, so that Eq.~(\ref{eq:Hb1I}) is not longer well approximated by Eq.~(\ref{eq:Hn}), and thus, a correct description requires its nonlinear counterpart, as described in~\ref{ss:BLD}. }}
\label{fig_scheme}
\end{figure}

The starting point of our theoretical framework consists in the consideration of a two-level system, described by the usual spin-${1}/{2}$ Pauli matrices $\vec{\sigma}=\left(\sigma_x,\sigma_y,\sigma_z\right)$ subject to rotations, coupled to a bosonic mode, in turn described by means of the annihilation and creation operators $a$ and $\adag$. The total Hamiltonian of the system can be written as
\begin{align}\label{eq:HG}
H_{\rm G}=H_{\rm spin}+H_{\rm boson}+H_{\rm int},
\end{align}
where the first and second terms comprise operators acting solely on the spin and the mode, respectively, thus reading (we assume units such that $\hbar=1$ throughout the paper)
\begin{align}\label{eq:HGspin}
  &H_{\rm spin}=\frac{\delta_0}{2}\sigma_x+\sum_{j=0}^{n_d} \frac{\Omega_j}{2}\left\{\cos \Delta_jt \ \sigma_z+\sin\Delta_jt \ \sigma_y \right\}, \\ \label{eq:HGboson}
  &H_{\rm boson}=\nu\adaga.
\end{align}
The third term corresponds to the spin-boson interaction. In particular, we assume that $H_{\rm int}$ contains only linear spin-boson exchange terms. Without loss of generality, such interaction term can be written as 
\begin{align}\label{eq:Hint}
H_{\rm int}=i\frac{\eta\nu}{2}\sigma_x(a-\adag),
\end{align}
where $\eta$ is a real and dimensionless parameter, 
and the frequencies $\Delta_j$ and $\delta_0$ are related as $\Delta_j=\delta_j-\delta_0$. Finally, $n_d$ is the total number of different drivings with amplitude $\Omega_j$ applied to the spin. Note that $\Delta_0=0$ by definition so that the first term in the sum simply provides the free energy term $\Omega_0\sigma_z/2$. For $n_d=0$ (or $\Omega_{j>0}=0$), $H_{\rm G}$ reduces thus to a standard spin system with frequency splitting $\Omega_0$ and bias parameter $\delta_0$, while the frequency of the bosonic mode is given by $\nu$ (see Fig.~\ref{fig_scheme}{\it (a)} for a schematic representation). Indeed, if $\delta_{j}=0\ \forall j$, the previous model takes a more recognizable form, namely that of a standard QRM. Therefore, the Hamiltonian $H_{\rm G}$ corresponds to a generalization of such model, including spin drivings. 
It is worth mentioning that the dimensionless parameter $\eta$ gives account for the coupling regime: while for $0\lesssim |\eta|/2\lesssim 0.1$ counter-rotating terms may be neglected, 
for $0.1\lesssim |\eta|/2\lesssim 1$ and $|\eta|/2\gtrsim 1$ one finds the so-called ultra-strong~\cite{Ciuti:05,Anappara:09,Forn:17,Kockum:18} and deep-strong coupling regimes~\cite{Casanova:10}, respectively (cf. Ref.~\cite{Rossatto:17} for a spectral classification of these regimes in the standard QRM). Note that we have not included the so-called $A^2$ term in $H_{\rm G}$ which may have a considerable impact in a cavity QED realization of $H_{\rm G}$~\cite{Rzazewski:75,Nataf:10,Vukics:14}. However, as we do not consider here a specific setup, we neglect such term while we refer to Appendix~\ref{app:a2} for a discussion on this issue.

In addition, we consider that the system undergoes dissipation due to the interaction with an environment~\cite{Breuer}, whose dynamics can be cast into the master equation
\begin{align}\label{eq:rhog}
\dot{\rho}_{\rm G}=-i\left[H_{\rm G},\rho_{\rm G}\right]+\mathcal{L}[\rho_{\rm G}],
  \end{align}
where $\mathcal{L}[\cdot]$ describes the non-unitary (dissipative) part of the dynamics. Moreover, we will assume that the superoperator $\mathcal{L}[\cdot]$ can be written in a diagonal Lindbladian form~\cite{Breuer}
\begin{align}\label{eq:Dlind}
\mathcal{L}[\rho_{\rm G}]{=}\!\sum_{k}\gamma_kD_{k}[\rho_{\rm G}]{=}\!\sum_k\!\gamma_k\left(\!F_k\rho_{\rm G} F^\dagger_k{-}\frac{1}{2}\!\!\left[F_k^{\dagger}F_k,\rho_{\rm G}\right]_+\!\right)\!,
  \end{align}
where $\left[\cdot,\cdot\right]_+$ stands for an anti-commutator and $F_k$ denotes the $k^\text{th}$ jump operator with rate $\gamma_k$ and dissipator $D_k[\cdot]$.

Our main goal now is to bring $H_{\rm G}$ into the form of a  model that involves multiphoton exchanges, i.e. a model  containing interaction terms akin to $\sigma^{\pm}a^{n}$ or $\sigma^{\pm}{\adag}^{n}$ , denoted here by $H_{\rm n}$. To achieve such goal we first transform $H_{\rm G}$ into $H_{a}$ by moving to a rotating frame. In particular, we define $H_{a}=H_{a,0}+H_{a,1}$ with $H_{a,0}=-\delta_0\sigma_x/2$ such that $H_{\rm G}\equiv H_{a,1}^I$, i.e., $H_{a,1}$ in the interaction picture of $H_{a,0}$. Then we perform a unitary transformation of the resulting Hamiltonian, $T^{\dagger}(i\eta/2)H_aT(i\eta/2)$, with  $T(\alpha)$ a spin-dependent displacement operator. In the spin basis, we have 
\begin{align}\label{eq:TSI}
  T(\alpha)=\frac{1}{\sqrt{2}}\left(
  \begin{matrix} \mathcal{D}^{\dagger}(\alpha) & \mathcal{D}(\alpha) \\ -\mathcal{D}^{\dagger}(\alpha) & \mathcal{D}(\alpha) \end{matrix} \right),
\end{align}
where $\mathcal{D}(\alpha)=e^{\alpha a^{\dagger}-\alpha^*a}$ is the displacement operator of amplitude $\alpha$~\cite{Scully}. Finally,  we move to an additional interaction picture with respect to $H_{b,0}=(\nu-\tilde{\nu})\adaga-\tilde{\omega}\sigma_z/2$.  Note that the transformation $T(\alpha)$ has been proposed in Ref.~\cite{MoyaCessa:16} to attain a fast trapped-ion implementation of the quantum Rabi model. In order to ease the notation, in the following we will use $T\equiv T(i\eta/2)$, unless otherwise specified, as  well as $U_{x}\equiv U_{x}(t,t_0)=\mathcal{T} e^{-i\int_{t_0}^tds H_x(s)}$ to denote the time-evolution propagator of the Hamiltonian $H_x$ ($\mathcal{T}$ accounts for time ordering), while the superscript $I$ stands for operators in the interaction picture.

As mentioned above, we start by moving to a rotating frame with respect to $H_{a,0}=-\delta_0\sigma_x/2$, such that $H_{\rm G}\equiv H_{a,1}^I\equiv U_{a,0}^{\dagger}(H_{a}-H_{a,0})U_{a,0}$, or equivalently, $H_a=H_{a,0}+U_{a,0}H_{\rm G}U_{a,0}^{\dagger}$. While $H_{\rm boson}$ and $H_{\rm int}$ commute with $U_{a,0}$, the time-dependent terms in $H_{\rm spin}$ do not. Recalling that $\Delta_j=\delta_j-\delta_0$, the transformed Hamiltonian $H_{a}$ becomes
\begin{align}\label{eq:Ha}
  H_a&=
  \nu\adaga+i\frac{\eta\nu}{2}\sigma_x(a-\adag)\\&+\frac{1}{2}\sum_{j=0}^{n_d}\Omega_j\left[\cos\delta_jt \ \sigma_z+\sin \delta_jt \ \sigma_y \right],
\end{align}
while the whole master  equation becomes 
\begin{align}
\dot{\rho}_{a}&=-i U_{a,0}\left[ H_{\rm G}+H_{a,0},\rho_G\right] U_{a,0}^{\dagger} +U_{a,0} \mathcal{L}[\rho_{\rm G}]U_{a,0}^{\dagger}\nonumber\\&=-i[H_{a},\rho_a]+U_{a,0} \mathcal{L}[U_{a,0}^{\dagger}\rho_{a}U_{a,0}]U_{a,0}^{\dagger}.
  \end{align}
We now perform a unitary transformation using the operator $T$ in Eq.~(\ref{eq:TSI}) with displacement amplitude $\alpha=i\eta/2$, such that
\begin{align}\label{eq:Hb}
H_b&\equiv T^{\dagger}(i\eta/2)H_aT(i\eta/2)\nonumber\\&=\nu\adaga+\frac{1}{2}\sum_{j=0}^{n_d}\Omega_j \left[\sigma^+e^{i\eta(a+\adag)}e^{-i\delta_jt}+{\rm H.c.} \right].
  \end{align}
Note that the previous Hamiltonian is similar to the one of an optically-trapped ion after performing the rotating-wave approximation and written in the rotating frame with respect to the free energy term of its  internal degree of freedom, driven by $n_d$ classical radiation fields with amplitude $\Omega_j$~\cite{Leibfried:03}. The dynamics in this new frame follows from
\begin{align}
T^{\dagger}\dot{\rho}_a T=&-i T^{\dagger}\left[H_a,\rho_a  \right]T\nonumber\\&+T^{\dagger}U_{a,0} \mathcal{L}[U_{a,0}^{\dagger}\rho_a U_{a,0}]U_{a,0}^{\dagger} T.
  \end{align}
As $\rho_b= T^{\dagger}\rho_aT$, using the definition given in Eq.~(\ref{eq:Hb}), we get
\begin{align}\label{eq:rhob}
\dot{\rho}_b=-i\left[H_b,\rho_b\right]+T^{\dagger}U_{a,0} \mathcal{L}[U_{a,0}^{\dagger}T\rho_bT^{\dagger} U_{a,0}]U_{a,0}^{\dagger} T.
\end{align}
It is worth noting that the transformation $H_{\rm G}\to H_b$ is unitary, and so is the one taking $\rho_G$ into $\rho_b$. Finally, from the Hamiltonian $H_b$ given in Eq.~(\ref{eq:Hb}) one can attain the desired multi-boson and spin couplings by moving to a suitable interaction picture. Indeed, defining $H_{b,0}=(\nu-\tilde{\nu})\adaga-\tilde{\omega}\sigma_z/2$, $H_{b,1}^I$ (with $H_{b,1}=H_{b}-H_{b,0}$) takes the following form
\begin{align}\label{eq:Hb1I}
H_{b,1}^I&\equiv U_{b,0}^{\dagger}H_{b,1}U_{b,0}=\tilde{\nu}\adaga+\frac{\tilde{\omega}}{2}\sigma_z\nonumber\\&+\sum_{j=0}^{n_d}\frac{\Omega_j}{2} \left[ \sigma^+e^{-i(\tilde{\omega}+\delta_j)t}e^{i\eta(a(t)+\adag(t))}+{\rm H.c.}\right],
\end{align}
where $a(t)=a e^{-i(\nu-\tilde\nu)t}$. By requiring the Lamb-Dicke condition $\eta\sqrt{\langle(a+\adag)^2 \rangle}{\ll}1$, one can expand the exponential term into power series, so that interaction terms like  $\sigma^+a^n$  or $\sigma^+(\adag)^n$ with $n\geq 1$ become resonant when selecting $\delta_j=\delta^{\pm}_n\equiv \pm n(\tilde\nu-\nu)-\tilde\omega$, while any other term will be rotating at frequency proportional to $\tilde\nu-\nu$. The previous condition is commonly known as Lamb-Dicke regime in the context of trapped ions~\cite{Leibfried:03}, while selecting frequencies $\delta_n^{\pm}$ corresponds to driving red- and blue-sideband processes. We will therefore refer to Lamb-Dicke regime to indicate such condition.  For small amplitudes, $\Omega_j\ll \nu$, one may neglect fast-oscillating terms by performing a rotating-wave approximation, i.e., preserving only those terms which are resonant. In this manner, we bring $H_{b,1}^I$ into the form of a spin-boson model with multi-boson interaction terms, denoted here by $H_{\rm n}$, so that $H_{b,1}^I\approx H_{\rm n}$.  Note however that the particular expression of $H_{\rm n}$ depends on the number $n_d$ of distinct drivings and their respective frequency $\delta_j$. The most general expression of the multi-boson model $H_{\rm n}$ within the Lamb-Dicke regime is
\begin{align}\label{eq:Hn}
  H_{\rm n}=&\tilde\nu\adaga+\frac{\tilde\omega}{2}\sigma_z+\sum_{m\in B}\left[\tilde{g}_me^{i\phi_m}\sigma^+(\adag)^m+{\rm H.c.}\right]\nonumber\\&+\sum_{n\in R}\left[\tilde{g}_ne^{i\phi_n}\sigma^+a^n+{\rm H.c.}\right],
  \end{align}
where the sets $R$ and $B$ encompass all the terms for which $\delta_j=\delta^{+}_n$ and $\delta_j=\delta_{m}^-$, respectively. We have set $\phi_n=n\pi/2$ and $\tilde{g}_n=\eta^n \Omega_{j,n}/(2\ n!)$, where $\Omega_{j,n}$ denotes the amplitude of the driving with frequency $\delta_{j}=\delta_{n}^{\pm}$. From Eq.~(\ref{eq:Hn}) we can see that, for a single term $\delta_0=\delta_n^+$ ($\delta_n^-$) with $n_d=0$, the resulting Hamiltonian becomes that of an $n$-boson (anti-)JCM. If an additional driving is introduced, one can bias the weights between rotating and counter-rotating terms in an $n$-photon QRM. Interestingly, by suitably adjusting the parameters  $\tilde\nu$ and $\tilde\omega$, different coupling regimes of these models are accessible, from weak ($\tilde\nu,\tilde\omega\gg \tilde g_n$) to deep-strong coupling ($\tilde{\nu}\lesssim \tilde{g}_n$). The latter regime however entails that longer evolution times under $H_{\rm G}$ are required to simulated $H_{\rm n}$, as $\tilde\nu\ll \nu$.

We would like to draw the attention to the Hamiltonian $H_{\rm n}$, which encompasses different models displaying fundamentally different physics. In Sec.~\ref{sec:res} we will analyze particular examples in which $H_{\rm n}$ reduces to the forms of  well-known models such as JCM, one-photon and two-photon QRMs~\cite{Travenec:12,Chen:12,Peng:13,Felicetti:15,Duan:16,Puebla:17pra}.

Finally, after moving to a rotating frame with respect to $H_{b,0}$ and performing the rotating wave approximation, Eq.~(\ref{eq:rhob})  becomes
\begin{align}\label{eq:rhonME}
\dot{\rho}_{\rm n}= -i[H_{\rm n},\rho_{\rm n}] + \tilde{\mathcal{L}}[\rho_{\rm n}],
  \end{align}
where now the transformed dissipative part is
\begin{align}\label{eq:DissMap}
\tilde{\mathcal{L}}[\cdot]=\Gamma \mathcal{L}[\Gamma^{\dagger}\cdot \Gamma]\Gamma^{\dagger} \quad {\rm with} \quad \Gamma=U_{b,0}^{\dagger}T^{\dagger}U_{a,0}.
  \end{align}
Hence $\Gamma$ is a unitary transformation that approximately maps the model $H_{\rm G}$ into $H_{\rm n}$. Recall that $U_{b,0}=e^{-i(t-t_0)((\nu-\tilde\nu)\adaga-\tilde\omega\sigma_z/2)}$, $T\equiv T(i\eta/2)$ as given in Eq.~(\ref{eq:TSI}) and $U_{a,0}=e^{i(t-t_0)\delta_0\sigma_x/2}$ with $t_0$ the initial time.  As a consequence, the structure of $\mathcal{L}[\rho_{\rm G}]$ (Eq.~(\ref{eq:Dlind})) is preserved, although with the replacement ${F}_k\to\tilde{F}_k=\Gamma F_k \Gamma^{\dagger}$.
The new jump operators $\tilde{F}_k$ are, in general, time-dependent  operators. Hence, although the transformed master equation resembles a Lindbladian one, the dynamics in general does not represent a semigroup~\cite{Breuer:16}. 

Nevertheless, the dynamics of an initial state $\rho_{\rm G}(t_0)$ evolving under Eq.~(\ref{eq:rhog}) approximately corresponds to the dynamics of $\rho_{\rm n}(t_0)=\Gamma^\dagger \rho_{\rm G}(t_0) \Gamma$ following Eq.~(\ref{eq:rhonME}), where the specific form of $H_{\rm n}$ crucially depends on the frequencies $\delta_j$, amplitudes $\Omega_j$ and number of terms $n_d$. We remark that the simulation of $H_{\rm n}$ starting from $H_{\rm G}$ holds to a very good approximation provided the previous conditions are satisfied, i.e., that $H_{b,1}^I$ can be well approximated by $H_{\rm n}$, thus
\begin{align}\label{eq:rhonrhoG}
  \rho_{\rm n}(t)\approx \Gamma \rho_{\rm G}(t) \Gamma^{\dagger}.
\end{align}
In other words, how well state $\rho_{\rm n}(t)$ can be realized from $\rho_{\rm G}(t)$ depends solely on how well the conditions for the application of the rotating wave approximation, which allows to neglect of fast-oscillating terms in Eq.~(\ref{eq:Hb1I}), are met. 
Eq.~(\ref{eq:rhonrhoG}) is the main result of the theoretical framework illustrated in this Section.

\subsection{Going beyond the Lamb-Dicke regime}\label{ss:BLD}
If the Lamb-Dicke condition $\eta\sqrt{\langle(a+\adag)^2 \rangle}{\ll}1$ is not satisfied, the exponential term in Eq.~(\ref{eq:Hb1I}) can not be expanded as carried out previously to achieve Eq.~(\ref{eq:Hn}). It is however still possible to write down a Hamiltonian after a suitable rotating wave approximation (beyond the Lamb-Dicke regime), resulting in nonlinear spin-boson terms~\cite{MatosFilho:94,Vogel:95,MatosFilho:96,Cheng:18,Krumm:18}. 
In this regime, the Hamiltonian $H_{b,1}^I$ is better approximated by the nonlinear counterpart of $H_{\rm n}$, i.e.
\begin{align}\label{eq:Hneta}
H_{b,1}^I\approx H_{\rm n,\eta}&=\tilde\nu\adaga+\frac{\tilde\omega}{2}\sigma_z\nonumber\\&+\sum_{m\in B}\frac{\Omega_{j,m}}{2} \left[\sigma^+(\adag)^nf_m(\adaga)+{\rm H.c.}\right]\nonumber\\&+\sum_{n\in R}\frac{\Omega_{j,n}}{2} \left[\sigma^+f_n(\adaga)a^n+{\rm H.c.}\right],
\end{align}
where we have introduced the operator $f_n(\adaga)$ 
\begin{align}\label{eq:fn}
  f_n(\adaga)=e^{-\eta^2/2}\sum_{l=0}^{\infty}\frac{(i\eta)^{2l+n}}{l!(l+n)!}(\adag)^la^l.
\end{align}
We stress that, although the previous model in Eq.~(\ref{eq:Hneta}) certainly contains a nonlinear spin-boson coupling (for $n>1$), there is yet another source of nonlinearity originated from the function $f_n(\adaga)$. Indeed, the transition rates between states $\ket{e,m}$ and $\ket{g,m+n}$ are effectively reduced depending on the value of $f_n(\adaga)$, and thus on that of $\eta$, which can vary significantly for varying Fock number [cf. Fig.~\ref{fig_scheme}{\it (c)}]~\cite{Vogel:95}. For a single term, that is in a nonlinear JCM, such feature hinders the appearance  of the hallmark of the standard JCM, namely, collapses and revivals of quantum population. Moreover, $f_n(\adaga)$ may vanish for a certain Fock state, thus exhibiting a blockade of the propagation of quantum amplitudes across the Hilbert space~\cite{Cheng:18}. Although we will refer to nonlinear models whenever $f_n(\adaga)$ has been taken into account, for the sake of clarity we indicate it  by introducing a subscript $\eta$ to the Hamiltonian [as in Eq.~(\ref{eq:Hneta})].

If one however finds itself within the Lamb-Dicke regime, then the previous Hamiltonian takes the form given in Eq.~(\ref{eq:Hn}), i.e., $H_{\rm n,\eta}\approx H_{\rm n}$ since $f_n(\adaga)$ becomes constant (see Fig.~\ref{fig_scheme}{\it (c)}). Note that for $\eta\sqrt{\langle(a+\adag)^2\rangle}\ll 1$, we have $f_n(\adaga)\approx (i\eta)^n/n!$ which, together with the amplitude $\Omega_{j,n}/2$, leads to the coupling given in Eq.~(\ref{eq:Hn}), $\tilde{g}_ne^{i\phi_n}$. Beyond their mathematical interest, these models display a number of interesting features with potential application in bosonic mode cooling~\cite{Morigi:99} or in dissipative state preparation of Fock states and nonlinear coherent states~\cite{MatosFilho:96,MatosFilho:96a,Cheng:18}. We will come back to these models in Sec.~\ref{sec:res}, in particular in~\ref{ss:naJCM} where we provide numerical simulations to illustrate their simulation using $H_{\rm G}$.


\section{Examples and numerical results}\label{sec:res}
In this Section we provide specific examples of the theoretical framework presented in Sec.~\ref{sec:theory}. We start by showing how typical jump operators transform under the map $\Gamma$ [cf.  Eq.~(\ref{eq:DissMap})]. Then, in~\ref{ss:Hn}, we show different examples of how the state satisfying Eq.~(\ref{eq:rhonME}) can be obtained from $\rho_{\rm G}$ obeying Eq.~(\ref{eq:rhog}), supported by numerical simulations and computing state fidelities between the ideal $\rho_{\rm n}(t)$ and its reconstructed version from $\rho_{\rm G}(t)$.

\subsection{Transformed jump operators}\label{ss:jump}
As we have explained in Sec.~\ref{sec:theory}, the jump operators $F_k$ affecting the Hamiltonian $H_{\rm G}$ map into $\tilde{F}_k=\Gamma F_k \Gamma^{\dagger}$. In the following we show the transformation for customary jump operators in quantum optics, namely, $\sigma_z$, $\sigma^{\pm}$ as well as $a$, $\adag$ and $\adaga$, which correspond to spin dephasing, spontaneous emission and absorption, boson leakage and heating, and boson dephasing, respectively~\cite{Breuer}.  In addition, we comment on what would be required to engineer a desired dissipative process $\tilde{F}_k$. One can calculate easily the transformed jump operators as explained in Sec.~\ref{sec:theory}, i.e., $\tilde F_k=\Gamma F_k\Gamma^{\dagger}$ where $\Gamma=U_{b,0}^\dagger T^{\dagger} U_{a,0}$, $U_{a,0}=e^{it\delta_0\sigma_x/2}$, $U_{b,0}=e^{-it((\nu-\tilde\nu)\adaga-\tilde\omega\sigma_z/2)}$ with $t_0=0$, and $T\equiv T(i\eta/2)$ as defined in Eq.~(\ref{eq:TSI}). As an example, in Appendix~\ref{app:jump} we provide the full derivation of the transformation of $F$ into $\tilde{F}$ for spontaneous emission and absorption. 

It is worth mentioning that, as reported in Ref.~\cite{Beaudoin:11}, considering independent  decoherence processes acting either on the spin or on the bosonic mode may become a crude approximation as their coupling enters in the ultra-strong regime [i.e. for $\eta/2\gtrsim 0.1$, cf. Eq.~(\ref{eq:Hint})]. It is then convenient to move to a suitable dressed basis, where the  relevant degrees of freedom are mixed up, and where one can adequately describe distinct dissipative processes~\cite{Beaudoin:11}. However, even in the ultra-strong coupling regime the differences in the steady-state populations between both approaches are small, in the order of $10^{-2}$ for $\left<\adaga\right>$~\cite{Beaudoin:11}. Thus, considering independent channels of dissipation may be considered as a reasonable approximation even for $\eta/2\gtrsim 0.1$. Although a dressed basis treatment lies outside the scope of this work, we provide a discussion in Appendix~\ref{app:dressed} on how the results vary when this is considered instead.

\textit{Spin dephasing.}--- The jump operator associated with spin dephasing reads $F_{\rm sd}=\sigma_z$, whose rate is given by $\gamma_{\rm sd}$. One thus obtains
\begin{align}\label{eq:sigmaz}
\Gamma \sigma_z\Gamma^{\dagger} = \mathcal{D}(t)e^{-it(\delta_0+\tilde\omega)}\sigma^++{\rm H.c.},
\end{align}
with the time-dependent displacement operator $\mathcal{D}(t)\equiv\mathcal{D}\left(i\eta e^{i(\nu-\tilde\nu)t}\right) =e^{i\eta(a(t)+\adag(t))}$.
Introducing the previous expression in the dissipator, we find
\begin{align}\label{eq:Dtildesd}
\tilde{D}_{\rm sd}[\rho]=&-\rho+\mathcal{D}(t)\sigma^+\rho \sigma^- \mathcal{D}^{\dagger}(t) + \mathcal{D}^{\dagger}(t)\sigma^-\rho \sigma^+ \mathcal{D}(t)\nonumber\\&+\left(\mathcal{D}(t)\sigma^+\rho \sigma^+ \mathcal{D}(t)e^{-2it(\delta_0+\tilde\omega)}+{\rm H.c.}\right).
  \end{align}
We thus observe that spin dephasing produces decoherence in the transformed frame by mixing spin and bosonic degrees of freedom. Indeed, neglecting fast oscillating terms, the previous dissipator contains in general nonlinear jump operators of the form $\sigma^{\pm}a^{n}$, $\sigma^{\pm}(\adag)^n$ and $\sigma^{\pm}(\adag)^na^n$ at order $\eta^n$.  We highlight that, although Eq.~(\ref{eq:sigmaz}) may seem to indicate that the dissipative terms  $\sigma^+ a^{n}$ or $\sigma^{-}(\adag)^n$ can be tuned by properly adjusting the frequency $\delta_0$ (as carried out to attain Eq.~(\ref{eq:Hn})), a correct description demands taking into account all the resonant terms appearing in Eq.~(\ref{eq:Dtildesd}) and not only those in Eq.~(\ref{eq:Hn}). However, within the Lamb-Dicke regime, it is still possible to approximate $\tilde{D}_{\rm sd}[\rho]$ by a simple expression. Indeed, for $\gamma_{\rm sd}\ll \nu$, one may consider only the zero-order term in $\eta$, so that
\begin{align}\label{eq:Dsdapprox}
  \tilde{D}_{\rm sd}[\rho]\approx \sigma_x\rho\sigma_x-\rho=D_{\sigma_x}[\rho].
  \end{align}
As the Lamb-Dicke condition breaks down, the previous approximation no longer holds, thus demanding the inclusion of the terms of Eq.~(\ref{eq:Dtildesd}). We refer to Appendix~\ref{app:diss} for numerical results in which nonlinear jump operators are crucial to correctly reproduce the targeted dissipative dynamics.

\textit{Spontaneous emission and absorption.}--- The jump operators associated to spontaneous emission and absorption processes at rates $\gamma_{\rm se}$ and $\gamma_{\rm sa}$ are $F_{\rm se}=\sigma^{-}$ and $F_{\rm sa}=\sigma^+$, respectively. Their transformed forms is [cf. Appendix~\ref{app:jump} for the derivation of such expressions]
\begin{equation}\label{eq:sigmapm}
\Gamma\sigma^{\pm}\Gamma^{\dagger}=\frac{1}{2}\left(-\sigma_z\pm \mathcal{D}(t)e^{-it(\delta_0+\tilde{\omega})}\sigma^{+}\mp {\rm H.c.} \right).
  \end{equation}
Hence, these processes lead into spin dephasing in the transformed picture, as well as mixed decoherence on the spin and bosonic degree of freedom, as $\mathcal{D}(t)e^{-it(\delta_0+\tilde\omega)}\sigma^+$ comprises nonlinear operators of the form $a^{n}\sigma^+$ and $(a^{\dagger})^{n}\sigma^+$. Furthermore, we can already notice that if $F=\sigma_x$, its transformed form becomes particularly simple, $\Gamma \sigma_x \Gamma^{\dagger}=-\sigma_z$, while for $\sigma_y$  a more intricate expression is attained, $\Gamma \sigma_y \Gamma^{\dagger}={\rm Re}[e^{i\eta(a(t)+\adag(t))}e^{-it(\delta_0+\tilde\omega)}]\sigma_y +{\rm Im}[e^{i\eta(a(t)+\adag(t))}e^{-it(\delta_0+\tilde\omega)}]\sigma_x $.  As in the case of spin dephasing, provided that $\gamma_{\rm se,sa}\ll \nu$, it is possible to approximate Eq.~(\ref{eq:sigmapm}) within the Lamb-Dicke regime as
\begin{align}\label{eq:Dseapprox}
\tilde{D}_{\rm se,sa}[\rho]\approx \frac{1}{4}\left(D_{\rm sd}[\rho]+D_{\rm se}[\rho]+D_{\rm sa}[\rho]\right).
  \end{align}
In Appendix~\ref{app:diss} we provide numerical results when the previous approximation does not hold and higher-order terms in Eq.~(\ref{eq:sigmapm}) become crucial to correctly reproduce the targeted dissipative dynamics.

\textit{Boson leakage, heating and dephasing.}--- The transformed jump operators for these boson dissipative processes, with rates $\gamma_{\rm bl}$, $\gamma_{\rm bh}$ and $\gamma_{\rm bd}$, respectively, read
\begin{align}
  \Gamma a\Gamma^{\dagger}&=a e^{-it (\nu-\tilde{\nu})}-i\frac{\eta}{2}\sigma_z, \\
  \Gamma \adag \Gamma^{\dagger}&=\adag e^{it (\nu-\tilde{\nu})}+i\frac{\eta}{2}\sigma_z,\\
  \Gamma \adaga \Gamma^{\dagger}&=\adaga+\sigma_z\frac{\eta}{2}\left(ia^{-it(\nu-\tilde{\nu})}+{\rm H.c.} \right), 
\end{align}
up to constant factors. Therefore, provided $|\nu-\tilde\nu|\gg \eta\gamma_{\rm bl, bh}$ so that terms like $\sigma_z\rho a^{\dagger}$ can be neglected, boson leakage and heating remain in the transformed frame $H_{\rm n}$ without modifying their rate $\gamma_{\rm bl, bh}$, and add spin dephasing at the reduced rate $\gamma_{\rm bl, bh}\eta^2/4$. In particular, for boson leakage, the transformed dissipator reads as
\begin{align}
  \tilde D_{\rm bl}[\rho]&=a\rho \adag-\frac{1}{2}\left[\adaga,\rho \right]_++\frac{\eta^2}{4}\left(\sigma_z\rho\sigma_z-\rho \right)\nonumber\\&\quad+\frac{\eta}{2}\left( ia(t)\rho\sigma_z-i\sigma_z\rho a^{\dagger}(t)\right)\\ \label{eq:Dbl_approx}
  &\approx D_{\rm bl}[\rho]+\frac{\eta^2}{4}D_{\rm sd}[\rho].
  \end{align}
Analogous considerations hold for boson heating. Similarly, boson dephasing leads approximately into
\begin{equation}
\tilde{D}_{\rm bd}[\rho]\approx D_{\rm bd}[\rho]+\frac{\eta^2}{4}D_{a\sigma_z}[\rho]+\frac{\eta^2}{4}D_{\adag\sigma_z}[\rho],
\end{equation}
and thus, while boson dephasing remains in the simulated model, it also produced decoherence mixing spin and boson degrees of freedom, with dissipators characterized by jump operators $a\sigma_z$ and $\adag\sigma_z$.

\textit{Engineered channel of dissipation.}--- Besides the customary dissipation processes, let us consider the situation in which a specific dissipative channel with jump operator $\tilde{F}$ is addressed. The simulation of $\tilde{F}$ requires thus $\Gamma^{\dagger}\tilde{F}\Gamma$ to be implemented in $\mathcal{L}[\rho_{\rm G}]$ [cf. Eq.~(\ref{eq:rhog})].  For the sake of clarity, we provide an example which we will exploit later on: if one aims to simulate spontaneous emission, $\tilde{F}=\sigma^{-}$, a dissipative process with $F=\frac{1}{2}\mathcal{D}(i\eta)(\sigma_z-i\sigma_y)$ needs to be included in $\mathcal{L}[\rho_{\rm G}]$ (see Appendix~\ref{app:jump} for its derivation). Note that, although the resulting processes $\Gamma^{\dagger}\tilde{F}\Gamma$   may be challenging for their experimental implementation, one may still resort to approximations depending on the precise parameters, as aforementioned.

\begin{figure}
\centering
\includegraphics[width=0.9\linewidth,angle=-90]{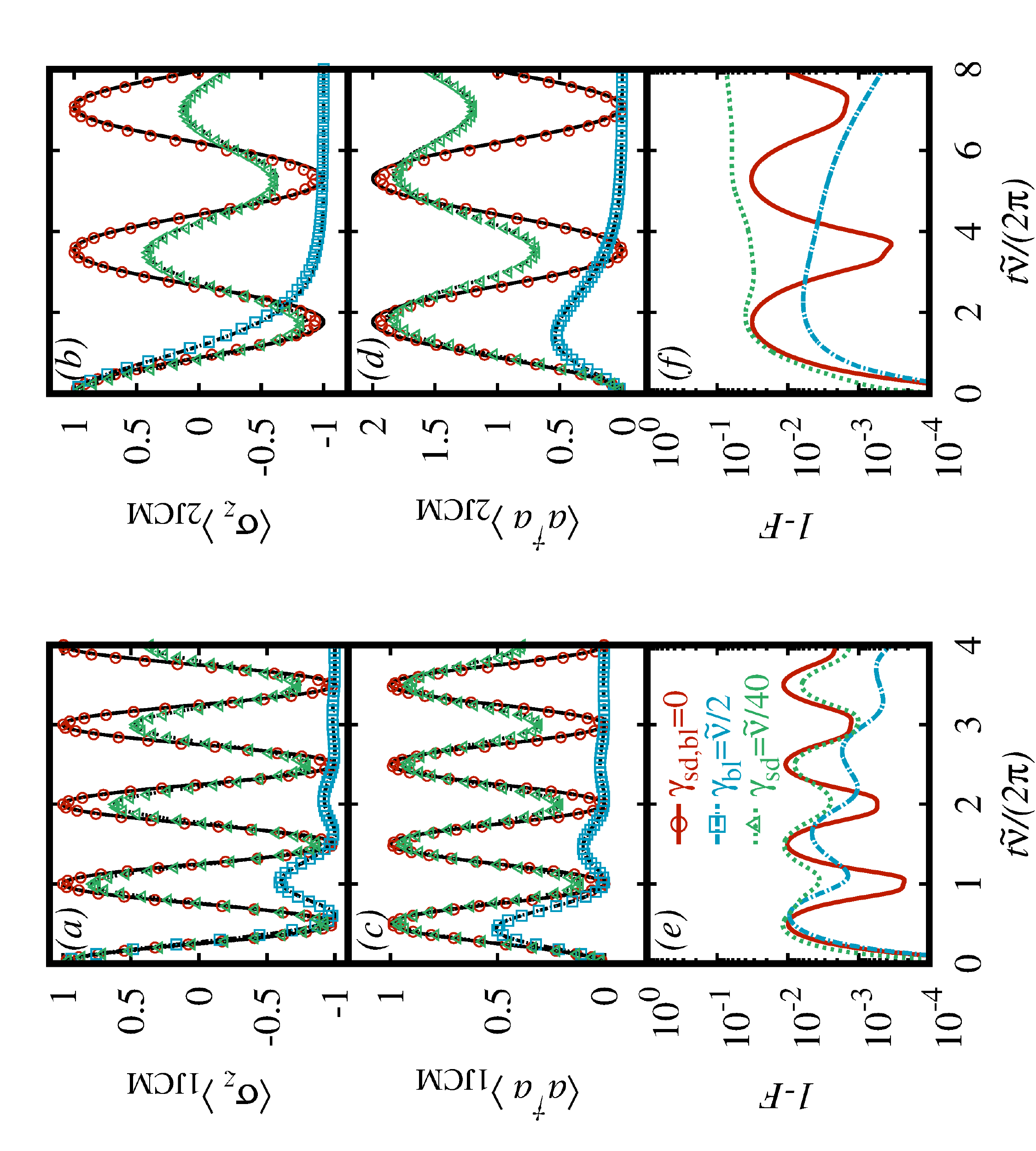}
\caption{\small{The dynamics induced by the nJCM embodied by the Hamiltonian $H_{\rm nJCM}$ in Eq.~(\ref{eq:nJCM}) is depicted by lines and compared to their simulation through $H_{\rm G}$ (shown by the points) as given in~\ref{ss:nJCM}. In the plots reported in the left column we have taken $n=1$, while $n=2$ has been used for the right column. In panels {\it (a)} and {\it (b)} [{\it (c)} and {\it (d)}] we show the evolution of $\left<\sigma_z\right>$ [$\left< \adaga\right>$],t while in {\it (e)} and {\it (f)} we plot the infidelity, $1-F(\rho_{\rm nJCM}(t),\Gamma \rho_{\rm G}(t)\Gamma^{\dagger})$ that quantifies the quality of the simulation obtained starting from $\rho_{\rm G}(t)$. The initial state $\ket{\psi(0) }_{\rm nJCM}=\ket{0}\ket{e}$ evolves  under $H_{\rm nJCM}$ without dissipation (solid red), and with dissipators corresponding to the transformed boson leakage (dashed blue) and spin dephasing (dotted green). The parameters used in the simulations are $\tilde\omega=n \tilde\nu$, $\tilde g_1=\tilde\nu/2$, $\tilde g_2=\tilde\nu/10$, $\Omega_0=\nu/100$, $\tilde\nu=\nu/2000$,  $\gamma_{\rm bl}=\tilde\nu/2$ and $\gamma_{\rm sd}=\tilde\nu/40$. 
}}
\label{fig_nJCM}
\end{figure}

\subsection{Examples of the theory: simulating $H_{\rm n}$ from $H_{\rm G}$}\label{ss:Hn}
In order to show the versatility and richness of the effects encompassed by the framework described in Sec.~\ref{sec:theory}, we specialize the general procedure to simulate a few interesting scenarios.

\subsubsection{$n$-boson JCM: $H_{\rm n}\rightarrow H_{\rm nJCM}$}\label{ss:nJCM}
Let us first consider  a simple case in which $n_d=0$ with $\delta_0=n(\tilde\nu-\nu)-\tilde\omega$. In this case, our starting Hamiltonian becomes time independent and takes the form of a generalized QRM
\begin{align}\label{eq:HgnJCM}
H_{\rm G}=\nu\adaga+\frac{\delta_0}{2}\sigma_x+\frac{\Omega_0}{2}\sigma_z+i\frac{\eta\nu}{2}\sigma_x(a-\adag),
  \end{align}
 where the bias parameter $\delta_0$ breaks explicitly the $Z_2$ symmetry. Although $H_{\rm G}$ above is in general nonintegrable, for $\delta_0=k \nu$ ($k\in\mathbb{Z}$), the model retrieves the  integrability of the standard QRM~\cite{Braak:11}.  The dynamics in this system obeys a master equation in Lindblad form, as introduced in Eqs.~(\ref{eq:rhog}) and~(\ref{eq:Dlind}). Interestingly, applying the map, this Hamiltonian  approximately corresponds to an $n$-boson JCM (that is, $H_{\rm n}$ adopts the form of $H_{\rm nJCM}$)
\begin{align}\label{eq:nJCM}
  H_{\rm nJCM}=&\tilde\nu\adaga+\frac{\tilde\omega}{2}\sigma_z+\tilde g_n\left[e^{i\phi_n}\sigma^+ a^n+{\rm H.c.}\right]
  \end{align}
with $\tilde g_n=\Omega_0\eta^n/(2 \ n!)$ and $\phi_n=n\pi/2$ [cf. Eq.~(\ref{eq:Hn})]. The forms taken by the initial state in each frame are related as $\rho_{\rm nJCM}(t_0)=T^{\dagger}\rho_{\rm G}(t_0)T$, while the state $\rho_{\rm nJCM}(t_0)$ evolves according to a master equation whose jump operators are in general  time dependent. 

The Hamiltonian $H_{\rm nJCM}$ results from the use of the rotating wave approximation, which requires both $\Omega_0\ll \nu$ and the Lamb-Dicke condition. Therefore, in order to realize an interacting nJCM ($\tilde{g}_n\sim \tilde{\nu}$), the parameters need to fulfill $\tilde\omega,\tilde\nu\ll \nu$.  In turn, this results in $\delta_0\approx -\nu$, which means that the integrability of the generalized QRM $H_{\rm G}$ in Eq.~(\ref{eq:HgnJCM}) is only weakly broken~\cite{Braak:11}. It is worth mentioning that, by taking $\delta_0=-n(\tilde\nu-\nu)-\tilde\omega$, one would attain an $n$-boson anti-JCM, an interaction term of the form $\sigma^- a^n +{\rm H.c.}$, as illustrated in Eq.~(\ref{eq:Hn}).  In Fig.~\ref{fig_nJCM}  we show the numerical results for a one- and two-boson JCM, including distinct channels for dissipation, to illustrate the good performance of simulating these models from a simple $H_{\rm G}$ (see~\ref{ss:num} for further details).

\subsubsection{$n$-boson QRM  $H_{\rm n}\rightarrow H_{\rm nQRM}$}\label{ss:nQRM}
\begin{figure}
\centering
\includegraphics[width=0.9\linewidth,angle=-90]{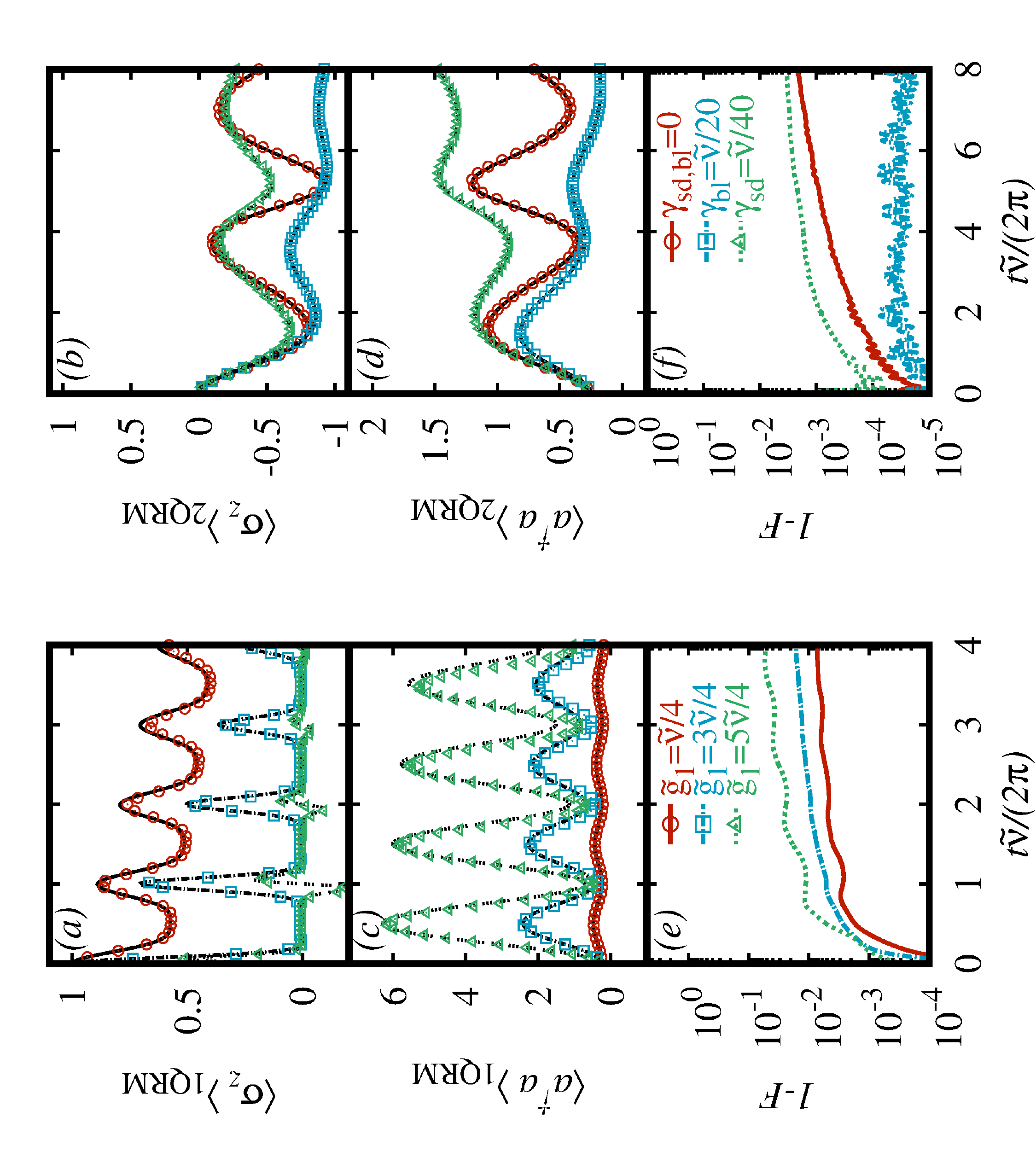}
\caption{\small{Dynamics of a nQRM, $H_{\rm nQRM}$ (see Eq.~(\ref{eq:nQRM})) for $n=1$ (left) and $n=2$ (right column), depicted by lines and their simulation using  $H_{\rm G}$ (points) as described in~\ref{ss:nQRM}. In panels {\it (a)} and {\it (b)} ({\it (c)} and {\it (d)}) we show the evolution of the expectation value $\left<\sigma_z\right>$ ($\left< \adaga\right>$),  while in {\it (e)} and {\it (f)} we plot the infidelity, $1-F(\rho_{\rm nQRM}(t),\Gamma \rho_{\rm G}(t)\Gamma^{\dagger})$ between the targeted state and its simulated counterpart. For the 1QRM (left column) we choose a pure initial state $\ket{\psi(0)}_{\rm 1QRM}=\ket{\alpha=1/2}\ket{e}$ where $\ket{\alpha}$ represents  a coherent state, while for the 2QRM the initial state reads $\rho_{\rm 2QRM}(0)=\rho_{b}^{\rm th}(0.25)\otimes \ket{+}\bra{+}$ where $\rho_{b}^{\rm th}(\left<\adaga\right>)$ is a thermal state with $\left<\adaga\right>$ average number of bosons. Different styles correspond to distinct coupling constants (left) or dissipation rates (right), as indicated in panels {\it (e)} and {\it (f)}, respectively. The parameters are $\tilde\nu=\nu/5000$ and  $\Omega=\nu/100$, while $\gamma_{\rm bl}=2\gamma_{\rm sd}=\tilde\nu/50$ and $\tilde\omega=0$ (1QRM) and $\tilde{g}_2=\tilde\nu/10$ with $\tilde\omega=2\tilde\nu$ (2QRM). See~\ref{ss:num} for further details.   }}
\label{fig_nQRM} 
\end{figure}

In this case, the Hamiltonian $H_{\rm G}$ contains two different time-dependent terms with amplitude $\Omega_1$ and frequencies $\Delta_1=\delta_1-\delta_0$ with $\delta_{0,1}=\pm n(\tilde\nu-\nu)-\tilde\omega$ [cf. Eq.~(\ref{eq:HGspin})]. After a suitable transformation, one obtains an $n$-photon QRM as
\begin{align}\label{eq:nQRM}
  H_{\rm nQRM}&=\tilde\nu\adaga+\frac{\tilde\omega}{2}\sigma_z \nonumber\\
&+\tilde g_n\left[e^{i\phi_n}\sigma^+ +e^{-i\phi_n}\sigma^-\right]\otimes \left[a^n+(\adag)^n \right],
\end{align}
that is, $H_{\rm n}$ adopts the form of $H_{\rm nQRM}$. This model holds provided that  $\Omega_0=\Omega_1$, while the parameters $\tilde{g}_n$ and $\phi_n$ are the same as the ones given for the nJCM and general $H_{\rm n}$ model. It is worth mentioning that, in general, one can simulate an anisotropic nQRM by simply tuning different amplitudes $\Omega_0$ and $\Omega_1$ in $H_{\rm G}$. The $n$-boson version of the quantum Rabi model (nQRM) has been mainly studied in its two-boson form ($n=2$), which is of importance in describing second-order processes in different quantum optics setups. The Hamiltonian $H_{\rm 2QRM}$ displays a remarkable feature: at $\tilde{g}_2=\tilde\nu/2$ and above a certain excitation energy, the eigenstates become those of a free particle and the spectrum turns into a continuum band~\cite{Travenec:12,Felicetti:15,Duan:16}.  Furthermore, upon the spectral collapse at $\tilde{g}_2=\tilde\nu/2$, the Hamiltonian becomes unbounded from below for $\tilde{g}_2>\tilde\nu/2$. For any nQRM with $n\geq 3$ the Hamiltonian  $H_{\rm nQRM}$ is unbounded both from below and above for any non-zero coupling $\tilde{g}_{n>2}\neq 0$~\cite{Lo:98}. 
We would like to remark that this allows to attain a strongly coupled QRM without the need of increasing the coupling in the original $H_{\rm G}$, i.e.,  from a weakly coupled spin-boson system. A similar result has been proposed in~\cite{Ballester:12}, although following a different strategy where ultra-strong and deep-strong coupling is achieved after a suitable interaction picture. In Fig.~\ref{fig_nQRM} we show numerical results  considering $n=1$ in different coupling regimes, and $n=2$ with $\tilde{g}_2<\tilde\nu/2$. The parameters are detailed in~\ref{ss:num}. The simulation of the unitary dynamics for $n=3$ (3QRM)  has been shown in Ref.~\cite{Casanova:18}.


\subsubsection{Beyond Lamb-Dicke: nonlinear $n$-boson anti-JCM $H_{\rm n,\eta}\rightarrow H_{\rm naJCM,\eta}$}\label{ss:naJCM}
As discussed in Sec.~\ref{ss:BLD}, the theoretical framework that we have presented can be exploited even beyond the Lamb-Dicke regime. Here we consider a nonlinear $n$-boson anti-JCM, whose nonlinear interaction terms have been already introduced in Eq.~(\ref{eq:Hneta}). Then, considering $\delta_0=-n(\tilde\nu-\nu)-\tilde\omega$ and $n_d=0$, the Hamiltonian $H_{\rm n,\eta}$ takes the form of $H_{\rm naJCM,\eta}$, which reads as
\begin{align}\label{eq:naJCMeta}
  H_{\rm naJCM,\eta}&=\frac{\tilde{\omega}}{2}\sigma_z+\tilde{\nu}\adaga\nonumber\\&+\frac{\Omega_0}{2}\left[\sigma^+(\adag)^nf_n(\adaga)+{\rm H.c.}\right].
  \end{align}
Again, we remark that although the previous models are certainly nonlinear in the sense that they involve $n$-boson and spin interaction terms, we make use here of the term nonlinear (subscript $\eta$) to indicates that the transition rates between $\ket{e,m}$ and $\ket{g,m+n}$ become nonlinear, and differ fundamentally from those of the standard (linear) one~\cite{Vogel:95}. The transition rates involve the function $f_n(\adaga)$ that can significantly modify these rates (see Eqs.~(\ref{eq:Hneta}) and~(\ref{eq:fn}), as well as Fig.~\ref{fig_scheme}({\it (c)}). Here we keep the convention used in previous works where these models have been dubbed \textit{nonlinear} although comprising linear, i.e. one-boson, spin-boson exchange interaction terms~\cite{Vogel:95,MatosFilho:96,Cheng:18}.
According to our theory, this nonlinear model can be indeed realized from a linear and time-independent Hamiltonian, Eq.~(\ref{eq:HgnJCM}). In order to observe a significant effect of the nonlinearity stemming from the latter one needs however a large $\eta$ parameter and/or bosonic population. The departure from the Lamb-Dicke regime, in which $f_n(\adaga)\approx e^{-\eta^2/2}(i\eta)^n/n!$,  can be essentially captured by plotting $\bra{m}f_n(\adaga)\ket{m}$ as a function of the Fock state $m$ and for different values of $\eta$. In Fig.~\ref{fig_scheme}{\it (c)} we plot  $\bra{m}f_{n}(\adaga)\ket{m}$ for $n=0$ and $1$ and two values of $\eta$, namely $\eta=0.05$ and $0.75$, which illustrate the considerable modification of the transition rates for the latter case. Indeed, one can find $\eta$ such that $f_n(\adaga)\ket{k}\approx 0$, and so the transition between states $\ket{e,k} \leftrightarrow \ket{g,k+n}$ is suppressed in a $H_{\rm nJCM,\eta}$. See Ref.~\cite{Cheng:18} for further a discussion and potential applications regarding this nonlinearity.

\begin{figure}
\centering
\includegraphics[width=0.9\linewidth,angle=-90]{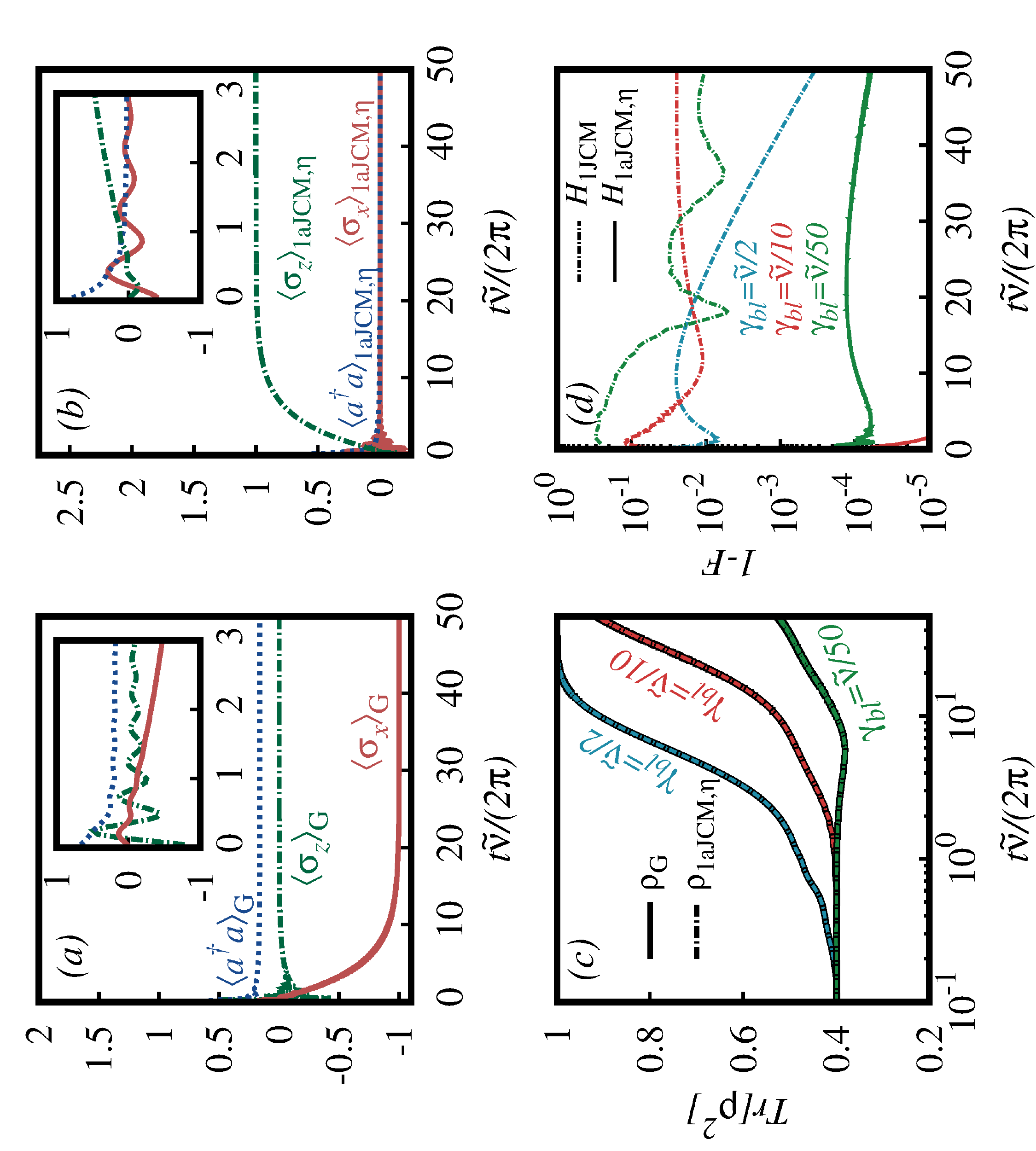}
\caption{\small{Dissipative dynamics towards the steady state in {\it (a)} $H_{\rm G}$ and {\it (b)} $H_{\rm 1aJCM,\eta}$ (see Eq.~(\ref{eq:nJCM})), where $H_{\rm G}$ is such that allows for the realization of the nonlinear 1JCM (see~\ref{ss:naJCM}), where $\tilde\omega=\tilde\nu= \Omega_0 f_1(0)/4$ (which would be equivalent to $\tilde{g}_1=\tilde\nu/2$ for $H_{\rm 1aJCM}$)  and $\tilde\nu=10^{-3}\nu$ with $\eta=0.8$. The initial state $\rho_{\rm G}(0)=\rho_b^{\rm th}(0.75)\otimes \ket{g}\bra{g}$ for $H_{\rm G}$ evolves under boson damping with rate $\gamma_{\rm bl}=\tilde\nu/2$ for {\it (a)}, and its transformed form for {\it (b)} (Eq.~(\ref{eq:Dbl_approx})) . The insets show the short-time dynamics. The panel {\it (c)} shows the time evolution of  the purity for $\rho_{\rm G}(t)$ (solid) and $\rho_{\rm 1aJCM,\eta}$ (dashed) while in {\it (d)} we plot the infidelity between $\Gamma \rho_{\rm G}(t)\Gamma^\dagger$ and $\rho_{\rm 1aJCM}$ (dashed lines) and $\rho_{\rm 1aJCM,\eta}$ (solid lines) for different rates, which for the latter may drop below our numerical precision $10^{-5}$.  See~\ref{ss:num} for further details and a discussion.}}
\label{fig_JCMBL}
\end{figure}

\begin{figure}
\centering
\includegraphics[width=0.9\linewidth,angle=-90]{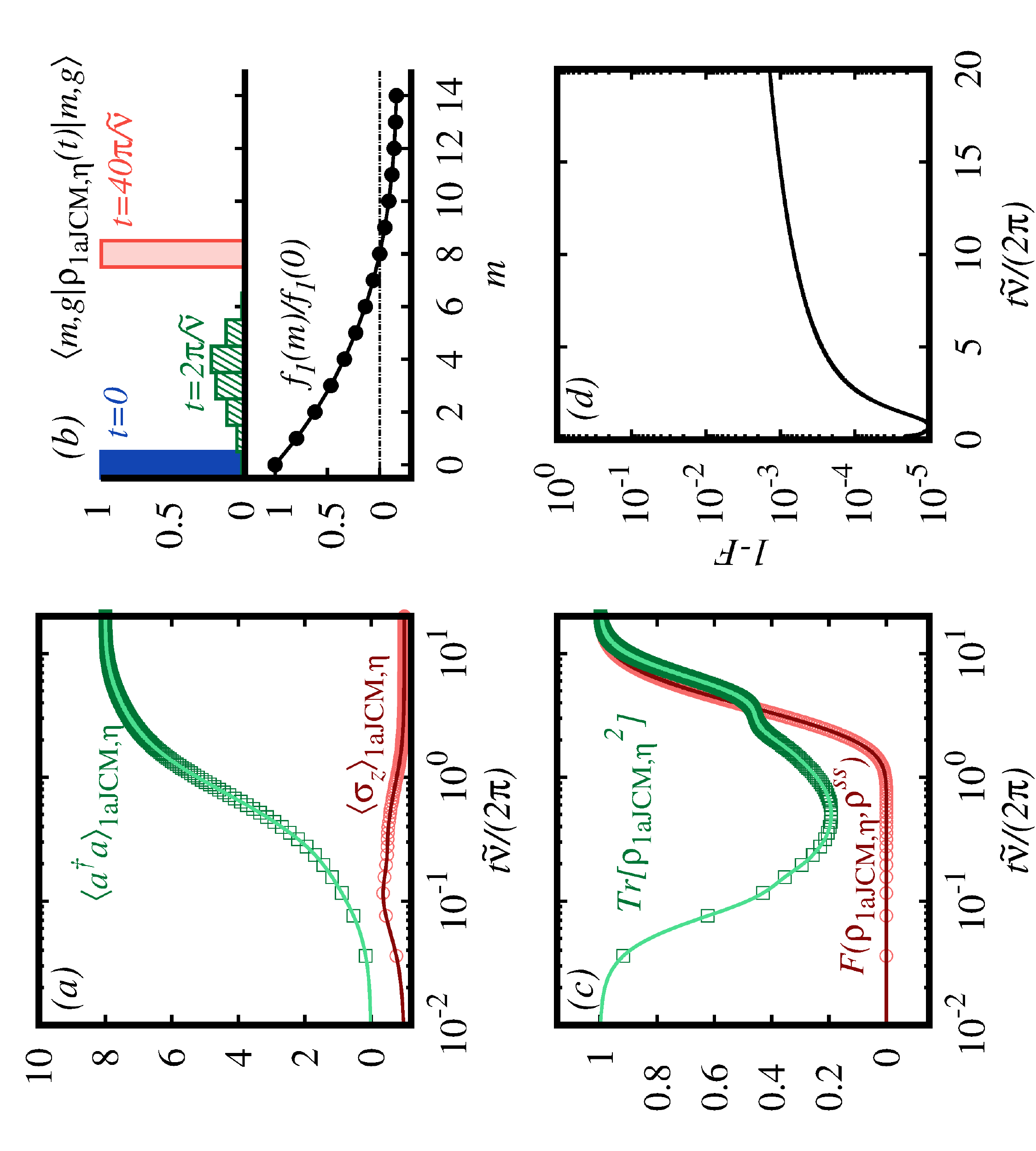}
\caption{\small{Dissipative preparation of a Fock state via $H_{\rm 1aJCM,\eta}$ and spontaneous emission, $\ket{\psi}^{ss}_{\rm 1aJCM,\eta}=\ket{m_s}\ket{g}$, which also corresponds to a pure state in the $H_{\rm G}$ counterpart but of the form $\ket{\psi}^{ss}_{\rm G}=\mathcal{D}(i\eta/2)\ket{m_s}\ket{+}$. In panel {\it (a)} we show the dynamics towards the steady state of $\rho_{\rm 1aJCM,\eta}$ (solid lines) and its simulation from $\rho_{\rm G}$ (points), starting from $\ket{\psi(0)}_{\rm 1aJCM,\eta}=\ket{0}\ket{g}$ and with $\tilde\omega=\tilde\nu=\Omega_0 f_1(0)$ (which would be equivalent to $\tilde{g}_1=2\tilde\nu$ for $H_{\rm 1aJCM}$) and $\gamma_{\rm se}=4\tilde\nu$. For $\eta\approx 0.64$, $f_1(m_s=8)$ vanishes, so that the state $\ket{m_s=8}\ket{g}$ becomes steady, see panel {\it (b)}. Panels {\it (c)} and {\it (d)} show the evolution of the purity (green), fidelity with respect to its steady and pure state $\rho^{ss}$ (red) for $\rho_{\rm 1aJCM,\eta}$ (lines) and $\rho_{\rm G}$ (points), and the infidelity between $\rho_{\rm 1aJCM,\eta}(t)$ and $\Gamma\rho_G(t)\Gamma^{\dagger}$.}}
\label{fig_JCMBL2}
\end{figure}

The nonlinear $n$-boson JCM $H_{\rm nJCM,\eta}$ can thus be realized only by having access to a generalized QRM, with a Hamiltonian of the form of $H_{\rm G}$ given in Eq.~(\ref{eq:HgnJCM}), with $\delta_0=n(\tilde\nu-\nu)-\tilde\omega$. Interestingly, nonlinear effects (as plotted in Fig.~\ref{fig_scheme}{\it (c)}) will become significant as $H_{\rm G}$ enters in the ultra-strong coupling regime, i.e. $\eta/2\gtrsim 0.1$. In a straightforward manner, one can realize different nonlinear models from a generalized quantum Rabi model in the ultra-strong coupling regime, such as a nonlinear nQRM~\cite{Cheng:18}. Note that for latter case one will proceed as explained in~\ref{ss:nQRM}, adding a spin driving into $H_{\rm G}$ and with a larger $\eta$ value such that ones goes beyond the Lamb-Dicke condition.   In Figs.~\ref{fig_JCMBL} and~\ref{fig_JCMBL2} we plot the results of the dynamics under $H_{\rm 1aJCM,\eta}$ and its reconstructed version using $H_{\rm G}$. See~\ref{ss:num} for further details regarding parameters and  dissipative processes.

\subsection{Numerical results}\label{ss:num}
In the following we present numerical results supporting the theoretical framework, specialized to the cases discussed previously, namely, where $H_{\rm G}$ allows for the realization of $H_{\rm nJCM}$, $H_{\rm nQRM}$ and $H_{\rm nJCM,\eta}$ with distinct dissipative processes. 

We start showing that the dynamics of an nJCM [Eq.~(\ref{eq:nJCM})] undergoing dissipative processes can be realized simply using a generalized QRM, whose form has been given in Eq.~(\ref{eq:HgnJCM}). In Fig.~\ref{fig_nJCM} we present numerical results of the simulation of the dynamics of a one- and two-boson JCM using $H_{\rm G}$ without dissipation, boson leakage and spin dephasing. The transformed dissipators for  spin dephasing and boson leakage have been approximated as in Eqs.~(\ref{eq:Dsdapprox}) and~(\ref{eq:Dbl_approx}), respectively, as explained in~\ref{ss:jump}. The parameters used for the simulations presented in Fig.~\ref{fig_nJCM} were $\tilde\omega=\tilde\nu=\tilde g_1/2$ (1JCM) and $\tilde\omega=2\tilde\nu=\tilde g_2/10$ (2JCM), while $\Omega_0=\nu/100$ and $\tilde\nu=\nu/2000$ which corresponds to $\eta=0.05$ (1JCM) and $\eta=0.14$ (2JCM), while the initial state considered for $H_{\rm nJCM}$ is $\ket{\psi(0)}_{\rm nJCM}=\ket{0}\ket{e}$. Hence, the initial state in the frame of $H_{\rm G}$ reads $\ket{\psi(0)}_{\rm G}=T^{\dagger}\ket{0}\ket{e}$.   The dissipation rates are $\gamma_{\rm bl}=\tilde\nu/2$ and $\gamma_{\rm sd}=\tilde\nu/40$. Note that, the approximation $\tilde{D}_{\rm sd}[\rho]\approx D_{\sigma_x}[\rho]$ does not work well in the 2JCM due to the larger value of the parameter $\eta$, as indicated by a larger infidelity in the latter case.  Therefore, our theory  allows us to reproduce the dynamics of these models in different dissipative and coupling regimes to a very good approximation, as demonstrated by the low infidelities [cf. Fig.~\ref{fig_nJCM}, panels {\it (e)} and {\it (f)}]. 

In the following, we consider the case where $H_{\rm G}$ allows us to simulate a dissipative $n$-boson quantum Rabi model, nQRM, whose Hamiltonian has been introduced in Eq.~(\ref{eq:nQRM}). We emphasize again that our starting point consists in a weakly coupled linear QRM, which allows strongly coupled spin-boson systems with $n$-boson exchange terms. In Fig.~\ref{fig_nQRM} we present numerical results in which a $H_{\rm 1QRM}$ is simulated in different regimes, even in the deep-strong coupling~\cite{Casanova:10} where $H_{\rm G}$ features a small coupling constant to bosonic frequency ratio $0.025$ (since $\eta = 0.05$ to attain $\tilde{g}_1/\tilde\nu=5/4$). The parameters of the simulated 1QRM are $\tilde\omega=0$ with coupling $\tilde{g}_1/\tilde\nu=1/4$, $3/4$ and $5/4$, while $\nu=5\times 10^3\tilde\nu$ and $\Omega_0=\Omega_1=50\tilde\nu$. The considered dissipative channels in $\mathcal{L}[\rho_{\rm G}]$ are boson leakage and spin dephasing with rates $\gamma_{\rm bl}=2\gamma_{\rm sd}=\tilde\nu/50$, while the chosen initial state is $\ket{\psi(0)}_{\rm 1QRM}=\ket{\alpha=1/2}\ket{e}$ where $\ket{\alpha}=\mathcal{D}(\alpha)\ket{0}$ denotes a coherent state, whose mean boson population amounts to $|\alpha|^2$. Notice that, as $\eta$ and the boson population are small, the approximation $\tilde{D}_{\rm sd}[\rho]\approx D_{\sigma_x}[\rho]$ is expected to hold. As shown in Fig.~\ref{fig_nQRM} {\it (a)} and {\it (c)}, as one increases the coupling $\tilde{g}_1/\tilde\nu$ one retrieves the main hallmark of the deep-strong coupling regime, namely, the structured collapses and revivals~\cite{Casanova:10}, which are damped here due to the dissipation. Although the agreement between the simulated dynamics and the targeted one is reasonably good [$F\gtrsim 0.9$ or $0.99$, depending on the case, cf. Fig.~\ref{fig_nQRM} {\it (e)}], we stress that tweaking the parameters may enhance further the attained fidelities.

In addition to the simulation of a 1QRM, we show that one can obtain good fidelities even for the realization of a 2QRM. For that, we select an initial state $\rho_{\rm 2QRM}(0)=\rho_{\rm b}^{\rm th}(\left<\adaga\right>)\otimes\rho_s$ where $\rho_{\rm b}^{\rm th}(\left<\adaga\right>)$ stands for a thermal state with average boson population $\left<\adaga\right>$, such that $\rho_b^{\rm th}(n)=\sum_{k=0}n^k(n+1)^{-k-1} \ket{k}\bra{k}$. For the results plotted in Fig.~\ref{fig_nQRM} we have considered $\rho_b^{\rm th}(0.25)$ while $\rho_s=\ket{+}\bra{+}$ where $\ket{\pm}=1/\sqrt{2}(\ket{e}\pm\ket{g})$. The parameters of the 2QRM are $\tilde\omega=2\tilde\nu$ and a coupling constant $\tilde{g}_2=\tilde\nu/10$, while for $H_{\rm G}$ we selected $\nu=5\times 10^3\tilde\nu=100\Omega_{0,1}$, which leads to $\eta\approx 0.09$. As in the previous cases, we consider boson losses and spin dephasing, where the latter can be still approximated as in Eq.~(\ref{eq:Dsdapprox}). The dynamics of this model is well reproduced in different cases, as shown in Fig.~\ref{fig_nQRM} {\it (b)}, {\it (d)} and {\it (f)}. Note that, when boson losses dominate the dynamics,  the state $\rho_{\rm 2QRM}(t)$ is pushed to the vacuum, $\ket{0}\bra{0}\otimes \ket{g}\bra{g}$, which becomes a the steady state. The attainment  of steady states upon different dissipative processes (and the correct functioning of their simulation) will be further inspected in the following, and illustrated in Figs.~\ref{fig_JCMBL} and~\ref{fig_JCMBL2}.

As commented in  Sec.~\ref{ss:BLD} and, for the specific case of the simulation of $H_{\rm naJCM,\eta}$, in Sec.~\ref{ss:naJCM}, a generalized QRM allows for the implementation of nonlinear, yet with $n$-boson exchange interactions, (anti-)JCMs. The main feature of this class of models is the strong dependence of transition rates between the states $\ket{e,m}\leftrightarrow\ket{g,m+n}$ for different $\ket{m}$ Fock states [cf. Fig.~\ref{fig_scheme} {\it (c)} and Sec.~\ref{ss:BLD}]. We illustrate the realization of a $H_{\rm 1aJCM,\eta}$ in Fig.~\ref{fig_JCMBL}, where the a steady state is achieved within the simulated evolution time. We take as initial state $\rho_{\rm G}(0)=\rho_{b}^{\rm th}(0.75)\otimes\rho_s$ with $\rho_s=\ket{g}\bra{g}$, $\eta=0.8$ and boson leakage with different rates, while the parameters are $\tilde\omega=\tilde\nu=\Omega_0f_1(0)/4$. We stress however that for such a large $\eta$ value, and thus large coupling constant in $H_{\rm G}$, considering independent channels of dissipation may be not longer a good approximation, and thus one may have to resort to a dressed-basis description of the dissipation (see Appendix~\ref{app:dressed}). Nevertheless, the difference in population may exhibit deviations in the order of $10^{-2}$ for the computed observables~\cite{Beaudoin:11}. In addition, note that the previous coupling constant would be equivalent to having $\tilde{g}_1=\tilde\nu/2$ in a standard $H_{\rm 1aJCM}$.  As shown in Fig.~\ref{fig_JCMBL}{\it (a)} and {\it (b)}, the steady state in $H_{\rm G}$ is accompanied by its transformed version in $H_{\rm 1aJCM,\eta}$. As the states $\rho_G(t)$ and $\rho_{\rm 1aJCM,\eta}$ are related through a unitary transformation, the purity ${\rm Tr}[\rho^2(t)]$ is expected to be equal if the simulation of this model works correctly. As we observe in Fig.~\ref{fig_JCMBL}{\it (c)}, this is indeed the case to a very good approximation. Note that for $\gamma_{\rm bl}=\tilde\nu/2$, the state $\rho_{\rm G}$ tends to a pure, yet steady, state, and so does $\rho_{\rm  1aJCM,\eta}$. As a matter of fact, as the dissipative dynamics for $H_{\rm naJCM,\eta}$ brings the states toward the  steady and pure state $\rho_{\rm nAJCM,\eta}^{ss}=\ket{0}\bra{0}\ket{e}\bra{e}$, it is easy to see that $\rho_{\rm G}^{ss}=\ket{\alpha=-i\eta/2}\ket{-}\bra{-}\bra{\alpha=-i\eta/2}$ is also pure, where $\ket{\alpha=-i\eta/2}=\mathcal{D}(-i\eta/2)\ket{0}$ is a coherent state containing $|\alpha|^2=\eta^2/4$ boson excitations. Finally, we comment that, due to the large coupling in $H_{\rm G}$ ($\eta=0.8$), the Lamb-Dicke condition breaks down and thus $H_{\rm 1aJCM}$ becomes a poor approximation of $H_{b,1}^I$ (Eq.~(\ref{eq:Hb1I})). In Fig.~\ref{fig_JCMBL}{\it (d)} we show the infidelity between $\Gamma\rho_{\rm G}(t)\Gamma^{\dagger}$ and the realized state $\rho_{\rm 1aJCM,\eta}(t)$ (solid lines) and $\rho_{\rm 1aJCM}(t)$ (dashed lines) for the same parameters. The significant difference among the fidelities, and the very low values for $1-F(\rho_{\rm 1aJCM,\eta}(t),\Gamma\rho_{\rm G}(t)\Gamma^{\dagger})\lesssim 10^{-4}$, pinpoints the correctness of the theory and the good realization of this nonlinear anti-Jaynes-Cummings model.

Finally, we provide a further example regarding the realization of a nonlinear anti-Jaynes-Cummings model when its nonlinearity is crucial, and thus $H_{\rm 1aJCM,\eta}$ differs fundamentally from its linear counterpart, $H_{\rm 1aJCM}$. For that we consider that spontaneous emission $\tilde{F}=\sigma^-$ can be implemented in $\mathcal{L}[\rho_{\rm n}]$, which corresponds to a dissipation channel with $F=\frac{1}{2}\mathcal{D}(i\eta)(\sigma_z-i\sigma_y)$ in $\mathcal{L}[\rho_{\rm G}]$ [cf. Sec.~\ref{ss:jump} and Appendix~\ref{app:jump} for its derivation]. As recently shown in Ref.~\cite{Cheng:18}, it is thus possible to tune $\eta$ such that $f_n(m_s)=0$ for a certain Fock state $\ket{m_s}$, and thus, spontaneous emission aids preparation of that precise Fock state, $\ket{m_s}\ket{g}$, since the transition to $\ket{m_s+n}\ket{e}$ is blocked. The numerical results regarding this situation are plotted in Fig.~\ref{fig_JCMBL2}, where the initial state $\ket{\psi(0)}_{\rm 1aJCM,\eta}=\ket{0}\ket{g}$ evolves towards $\ket{\psi}_{\rm 1aJCM,\eta}^{ss}=\ket{m_s}\ket{g}$ with $\eta=0.639$ such that $f_1(m_s=8)=0$ [Fig.~\ref{fig_JCMBL2} {\it (b)}], and with parameters $\tilde\omega=\tilde\nu=\Omega_0f_1(0)$ and dissipation rate $\gamma_{\rm se}=4\tilde\nu$. It is worth highlighting that the condition $\tilde\nu=\Omega_0f_1(0)$ would correspond to $\tilde\nu=2\tilde{g}_1$ in the standard $H_{\rm 1aJCM}$.  The dissipative preparation of the Fock state can be thus simulated by $H_{\rm G}$, where instead $\ket{\psi}_{\rm G}^{ss}=\mathcal{D}(i\eta/2)\ket{m_s}\ket{+}$. As in the previous case, the purity ${\rm Tr}[\rho^2(t)]$ matches for both states and the high fidelity, $F(\rho_{\rm 1aJCM,\eta}(t),\Gamma\rho_{\rm G}(t)\Gamma^{\dagger})\gtrsim 0.999$ [Fig.~\ref{fig_JCMBL2} {\it (c)} and {\it (d)}] indicate the good agreement between these models, where the parameters of $H_{\rm G}$ are $\nu=10^3\tilde\nu$ and $\Omega_0\approx \nu/130$. In addition, we include in Fig.~\ref{fig_JCMBL2}{\it (c)} the fidelity between the evolved state $\rho_{\rm G}(t)$ and the steady state $\ket{\psi}_{\rm G}^{ss}$, as well as  $\rho_{\rm 1aJCM,\eta}(t)$ and $\ket{\psi}_{\rm 1aJCM,\eta}^{ss}$, which evolves from $0$ to $1$ in a time $t\approx 40\pi/\tilde\nu$.

\section{Conclusions}\label{sec:conc}
We have shown how the dynamics of a dissipative quantum system comprising simultaneous exchanges of $n$-boson excitations with a spin can be realized exploiting a system involving only linear spin-boson coupling and simple spin rotations, i.e. without having access to the required $n$-boson  interacting terms. Moreover, we have demonstrate that spin-boson models with further nonlinear effects, such as the dependence of transition rates between Fock states on the actual Fock state number, can be accessed only by bringing a linear QRM into the ultra-strong coupling regime and with a spin bias.  Indeed, the simulation of these nonlinear and multiphoton spin-boson models can be realized in distinct regimes, ranging from weak to deep-strong coupling regimes. These models include  the well-known (anti-)JCMs and QRMs, both with and without nonlinearities, as well as the so-called two-boson QRM, among others. The developed theoretical framework not only unveils a deep connection between these models, but also offers the possibility for the simulation of nonlinear models previously constrained mainly to optical trapped-ion setups~\cite{Vogel:95,Cheng:18} and of multi-boson spin-boson interaction terms in platforms where those are otherwise unattainable, i.e., without relying on the developed theory.  

Here we have assumed that one has control onto a system described as a generalized QRM, i.e., spin and boson linearly coupled and with the ability of performing spin drivings, and it is where the simulation of a nonlinear $n$-boson spin-boson model is performed. The dissipative dynamics of the generalized QRM is assumed to be well described by a Lindblad term. In this manner, the jump operators then take a transformed form in the simulated model, which in general may become time dependent and involve also nonlinear spin-boson terms, thus mixing both degrees of freedom. For certain parameter regimes, however, distinct dissipative processes can be well approximated by standard dissipative channels.  We provide, in addition, a prescription of what jump operators would be required to be implemented in the simulator to achieve an arbitrary dissipative channel.  
We have then illustrated  the presented theoretical framework by showing  examples in which the dissipative dynamics of a generalized QRM corresponds to that of different nonlinear models with customary decoherence processes, such as spin and boson dephasing, spontaneous spin emission and absorption, and boson leakage and heating. The numerical simulations strongly support the theoretical results, indicating that the simulation holds to a very good approximation, as quantified in terms of the resulting high fidelities, for paradigmatic examples as the JCM and QRMs and their $2$-boson counterparts. These include the deep-strong coupling regime of the quantum Rabi model~\cite{Casanova:10}. Moreover, we also illustrate the simulation of a nonlinear anti-JCM, whose main trait consists in the blockade of propagation of quantum amplitudes along the Hilbert space~\cite{Cheng:18}.

Our results indicate that the dissipative dynamics of a generalized QRM approximately corresponds to different nonlinear models upon a suitable transformation, both of the coherent and dissipative parts.  Due to the ubiquity of a generalized QRM in a variety of quantum platforms and its relevance in different branches of modern science,  our results might open new avenues in the inspection of decoherence in different fundamental quantum systems and in their simulation.

\acknowledgements
R. P., O. H. and M. P. acknowledge the support by the SFI-DfE Investigator Programme (grant 15/IA/2864), and the H2020 Collaborative Project TEQ (Grant Agreement 766900). J. C. acknowledges support by the Juan de la Cierva grant IJCI-2016-29681. We also acknowledge funding from Spanish MINECO/FEDER FIS2015-69983-P and Basque Government IT986-16. This material is also based upon work supported by the U.S. Department of Energy, Office of Science, Office of Advance Scientific Computing Research (ASCR), Quantum Algorithm Teams (QAT) program under field work proposal number ERKJ333. J. C. and E. S. acknowledge support from the projects QMiCS (820505) and OpenSuperQ (820363) of the EU Flagship on Quantum Technologies. R. P. gratefully acknowledges the hospitality of J. C. and E. S. during his stay at the UPV/EHU.

\appendix

\section{Impact of an $A^2$-term in $H_{\rm G}$}\label{app:a2}
As mentioned in the main text, our starting point consists in considering a driven spin linearly coupled to a single bosonic mode whose Hamiltonian $H_{\rm G}$ is given in Eq.~(\ref{eq:HG}) and Eqs.~(\ref{eq:HGspin}),~(\ref{eq:HGboson}) and~(\ref{eq:Hint}). Although this description is wide since it applies to different setups, one may have to introduce an extra term $H_{A^2}$ when dealing, for example, with a cavity QED implementation as it can have a significant impact in the ultrastrong coupling regime~\cite{Rzazewski:75,Vukics:14,Nataf:10}. Note that this term stems from the potential vector of the cavity field. In the following lines we provide a discussion on how the results presented in the main text would be modified when this term is considered. In particular, when the $A^2$ term is included, our starting Hamiltonian becomes $H_{\rm G}=H_{\rm spin}+H_{\rm boson}+H_{\rm int}+H_{A^2}$,  where the latter term reads as
\begin{align}
H_{A^2}=D(a+\adag)^2.
\end{align}
Now, we consider an interaction $H_{\rm int}=\eta\nu/2 \sigma_x(a+\adag)$, which results from a trivial rotation of $a$ and $\adag$ in Eq.~(\ref{eq:Hint}). 
The previous Hamiltonian can be transformed back to the original form, i.e., without an $A^2$ term, by making a unitary transformation using the squeezing operator $\mathcal{S}[z]=e^{z/2((\adag)^2-a^2)}$ with $z\in \mathbb{R}$.  Then, calculating $\mathcal{S}^{\dagger}[z]H_{\rm G}\mathcal{S}[z]$ one can find the value $z_s$ that brings the previous expression into the form of $H_{\rm G}$ without $H_{A^2}$. Indeed, for $z_s=-1/4 \log(1+4D/\nu)$ the transformed Hamiltonian reads as
\begin{align}
\mathcal{S}^{\dagger}[z_s]H_{\rm G}\mathcal{S}[z_s]&=H_{\rm spin}+\frac{\tilde{\eta}\tilde{\nu}}{2}\sigma_x(a+\adag)\nonumber\\&+\tilde{\nu} e^{-2z_s}\adaga-\nu e^{-z_s}\sinh z_s,
  \end{align}
which is identical to the original Hamiltonian $H_{\rm G}$ up to constant values, see Eqs.~(\ref{eq:HGspin}),~(\ref{eq:HGboson}) and~(\ref{eq:Hint}),  with renormalized parameters, $\tilde{\eta}=\eta e^{3z_s}$ and $\tilde{\nu}=\nu e^{-2z_s}$. Hence, one can follow the rest of the theoretical derivation to find the nonlinear and multi-boson models as explained in Sec.~\ref{sec:theory}. Note however that the corresponding master equation describing the dynamics for $H_{\rm G}$ including the $H_{A^2}$ term may have to be transformed too according to $\mathcal{S}[z_s]\cdot \mathcal{S}^{\dagger}[z_s]$.

\section{Transformed jump operators for spontaneous emission and absorption}\label{app:jump}
As given in the main text, Eq.~(\ref{eq:sigmapm}), the jump operator associated with spontaneous absorption (emission) acting in the frame of $H_{\rm G}$, $F=\sigma^\pm$, transforms into the frame $H_{\rm n}$ as $\tilde{F}=\Gamma F \Gamma^{\dagger}$, where $\Gamma=U_{b,0}^{\dagger} T^{\dagger} U_{a,0}$. The time-evolution propagators are $U_{a,0}=e^{it\delta_0\sigma_x/2}$ and $U_{b,0}=e^{-it((\nu-\tilde\nu)\adaga-\tilde\omega\sigma_z)}$ considering $t_0=0$, and $T=2^{-1/2}\left[ \mathcal{D}(\alpha)(\ket{e}\bra{g}+\ket{g}\bra{g})+\mathcal{D}^{\dagger}(\alpha)(\ket{e}\bra{e}-\ket{g}\bra{e})\right]$ given in Eq.~(\ref{eq:TSI}). Thus, the transformed jump operator $\tilde{F}$ can be calculated as follows. We first need 
\begin{align}
  &U_{a,0}\ \sigma^\pm U_{a,0}^{\dagger}\nonumber\\&=\cos^2\left(\delta_0t/2 \right)\sigma^\pm+\sin^2\left(\delta_0t/2 \right)\sigma^\mp\mp\frac{i}{2}\sin(\delta_0t)\sigma_z,
\end{align}
and then, using the expressions
\begin{align}
  T^{\dagger}\sigma_zT&=\mathcal{D}^2(\alpha)\sigma^++\mathcal{D}^{\dagger 2}(\alpha)\sigma^-\\
  T^{\dagger}\sigma^{\pm}T&=\frac{1}{2}\left[ -\sigma_z\pm\mathcal{D}^2(\alpha)\sigma^+ \mp\mathcal{D}^{\dagger 2}(\alpha) \sigma^- \right]
\end{align}
we obtain
\begin{align}
T^{\dagger}U_{a,0}\ \sigma^+U_{a,0}^{\dagger}T=&-\frac{1}{2}\sigma_z+\frac{1}{2}\mathcal{D}^2(\alpha)e^{-it\delta_0}\sigma^+\nonumber\\&-\frac{1}{2}\mathcal{D}^{\dagger 2}(\alpha)e^{it\delta_0}\sigma^-,
\end{align}
so that finally we arrive to
\begin{align}
  &\Gamma \sigma^\pm \Gamma^{\dagger}=U_{b,0}^{\dagger}T^{\dagger}U_{a,0}\ \sigma^\pm U_{a,0}^{\dagger}TU_{b,0}=-\frac{1}{2}\sigma_z\nonumber\\&\pm\frac{1}{2}e^{it(\nu-\tilde{\nu})\adaga}\mathcal{D}(2\alpha)e^{-it(\nu-\tilde{\nu})\adaga}e^{-it(\delta_0+\tilde{\omega})}\sigma^+\nonumber\\&\mp\frac{1}{2}e^{it(\nu-\tilde{\nu})\adaga}\mathcal{D}^{\dagger}(2\alpha)e^{-it(\nu-\tilde{\nu})\adaga}e^{it(\delta_0+\tilde{\omega})}\sigma^-
\end{align}
which for $\alpha=i\eta/2$, $\mathcal{D}(i\eta)=e^{i\eta (\adag+a)}$, and therefore
\begin{align}\label{eqa:tildesp}
\Gamma \sigma^+\Gamma^{\dagger}=&-\frac{1}{2}\sigma_z\pm \frac{1}{2}e^{i\eta(a(t)+\adag(t))}e^{-it(\delta_0+\tilde{\omega})}\sigma^\pm\nonumber\\&\mp\frac{1}{2}e^{-i\eta(a(t)+\adag(t))}e^{it(\delta_0+\tilde{\omega})}\sigma^-
\end{align}
with $a(t)=ae^{-it(\nu-\tilde{\nu})}$. The previous expression corresponds to the Eq.~(\ref{eq:sigmapm}) given in the main text.

\begin{figure}
\centering
\includegraphics[width=0.9\linewidth,angle=-90]{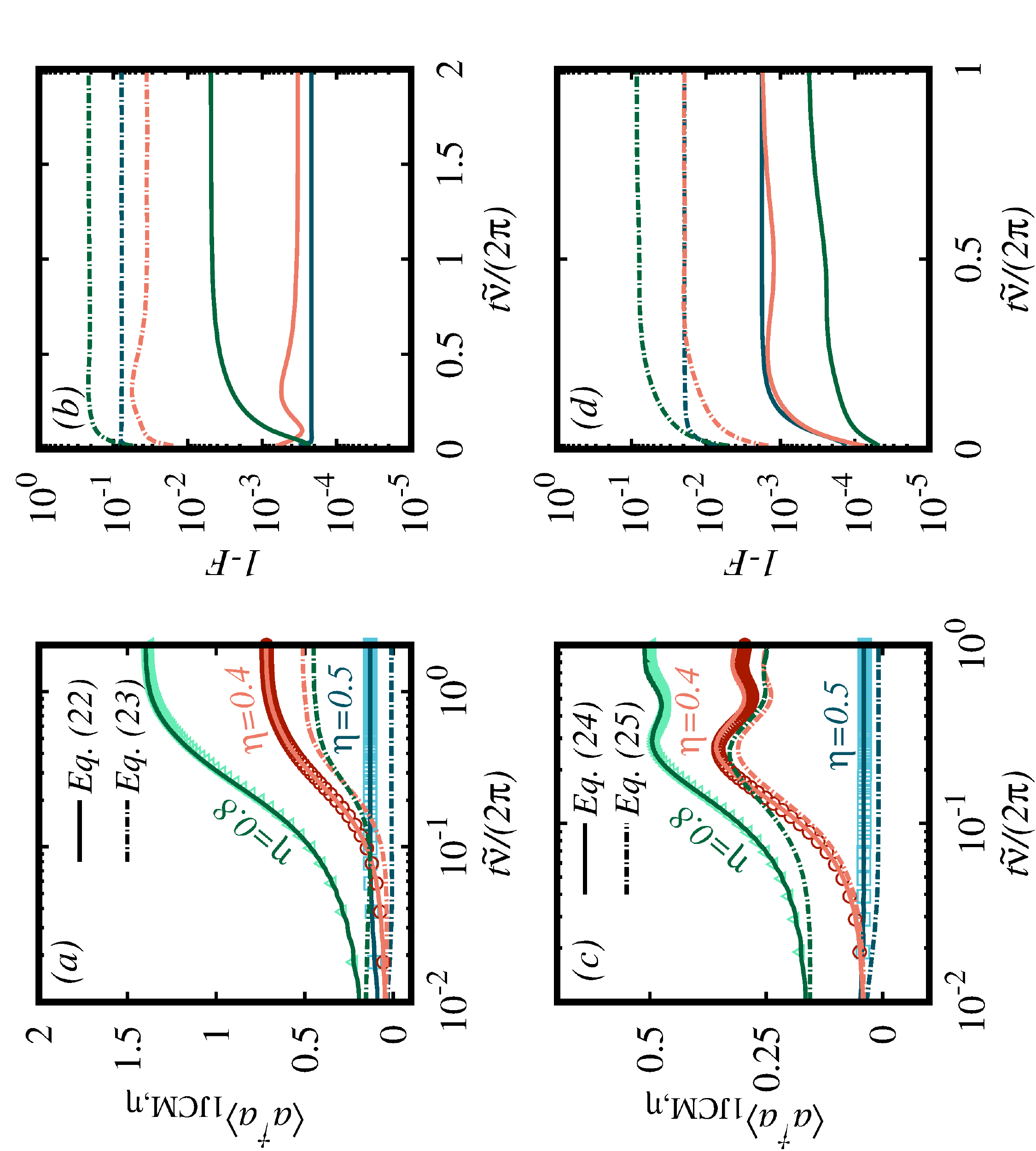}
\caption{\small{Dissipative dynamics when $\tilde{D}_{\rm sd}[\rho]$ (top) and $\tilde{D}_{\rm se}[\rho]$ (bottom) can not be approximated by the simple Eqs.~(\ref{eq:Dsdapprox}), and~(\ref{eq:Dseapprox}), respectively. In {\it (a)} we show the time evolution of $\left<\adaga \right>_{\rm 1JCM,\eta}$ (lines) and its reconstruction using $\rho_{\rm G}$ (points), when $\tilde{D}_{\rm sd}[\rho]$ is taken as in Eq.~(\ref{eq:Dtildesd}) (solid lines) or approximated as in Eq.~(\ref{eq:Dsdapprox}) (dashed lines), for different parameters, namely $\eta=0.5$ with $\gamma_{\rm sd}=\gamma_{\rm bl}/2=5\tilde\nu$ (blue) and $\eta=0.4$ and $0.8$ with $\gamma_{\rm sd}=2\gamma_{\rm bl}=\tilde\nu$ (red and green). The rest of the parameters are $\tilde\omega=\tilde\nu=f_1(0)\Omega_0/2$ with $\tilde\nu=10^{-2}\nu$, and initial state $\ket{\psi(0)}_{\rm G}=\ket{0}\ket{+}$. With the same format, in {\it (b)} we show the infidelity $1-F$ between the $\rho_{\rm G}$ and $\rho_{\rm 1JCM,\eta}$ considering $\tilde{D}_{\rm sd}[\rho]$ as in Eq.~(\ref{eq:Dtildesd}) or Eq.~(\ref{eq:Dsdapprox}). The same is plotted in the bottom panels, {\it (c)} and {\it (d)}, but with $\tilde{D}_{\rm se}[\rho]$ and $\gamma_{\rm se}$ instead of $\tilde{D}_{\rm sd}[\rho]$ and $\gamma_{\rm sd}$ and $\ket{\psi(0)}_{\rm G}=\ket{0}\ket{e}$.}}
\label{fig_SpinDephFull}
\end{figure}

In a straightforward manner, for $\tilde{F}=\sigma^-$ the corresponding jump operator in the frame of $H_{\rm G}$ follows from $F=\Gamma^{\dagger}\tilde{F}\Gamma$,
\begin{align}
  \Gamma^{\dagger}\tilde{F}\Gamma&=U_{a,0}^{\dagger}TU_{b,0}\sigma^- U_{b,0}^{\dagger}T^{\dagger}U_{a,0}\\
  &=e^{-it\tilde\omega}U_{a,0}^{\dagger}T\sigma^- T^{\dagger}U_{a,0}\\
  &=\frac{1}{2}\mathcal{D}(2\alpha)e^{-it\tilde\omega }U_{a,0}^{\dagger}\left[\sigma_z-i\sigma_y\right]U_{a,0}\\
  &=\frac{1}{2}\mathcal{D}(2\alpha)e^{-it(\tilde\omega+\delta_0)}\left[\sigma_z-i\sigma_y\right].
  \end{align}
Since $e^{-it(\tilde\omega+\delta_0)}$ is just a global phase, it can be dropped out, so a dissipative channel $F=\frac{1}{2}\mathcal{D}(i\eta)\left[\sigma_z-i\sigma_y\right]$ in $H_{\rm G}$ leads to spontaneous emission in $H_{\rm n}$.

\section{Dressed-basis treatment of dissipation}\label{app:dressed}
As acknowledged in Ref.~\cite{Beaudoin:11}, the master equation considering independent channels of dissipation leads into non-physical results as the coupling constant becomes comparable to the bosonic frequency. Instead, one needs to describe dissipation in the dressed basis of the spin-mode. In the case of a time-independent $H_{\rm G}=\sum_k E_k \ket{k}\bra{k}$ where $\ket{k}$ denotes here the $k$th eigenstate of the spin-boson system. Then, the correct master equation at zero temperature involving spin dephasing and boson leakage results in
\begin{align}
\mathcal{L}[\rho_{\rm G}]=D_A[\rho_{\rm G}]&+ \sum_{j,k\neq j}\gamma_{{\rm sd},jk}D_{B_{jk}}[\rho_{\rm G}]\nonumber \\&+\sum_{j,k>j}\gamma_{{\rm bl},jk} D_{B_{jk}}[\rho_{\rm G}],
\end{align}
with the operators $A=\sum_k \sqrt{\gamma_{\rm sd}(0)}\bra{k}\sigma_z\ket{k} \ket{k}\bra{k}$ and $B_{jk}=\ket{j}\bra{k}$ with $\gamma_{{\rm sd},jk}=\gamma_{\rm sd}(E_k-E_j) \left|\bra{j}\sigma_z\ket{k} \right|^2$ and $\gamma_{{\rm bl},jk}=\gamma_{\rm bl}(E_k-E_j) \left|\bra{j}(a+\adag)\ket{k} \right|^2$, where the rates are now evaluated at different frequencies. Performing the transformation described in Sec.~\ref{sec:theory}, one would find the correct master equation describing the dynamics of the simulated multi-boson and nonlinear models. In particular, one would have to transform the operators $A$ and $B_{jk}$ as $\Gamma A \Gamma^{\dagger}$ and  $\Gamma B_{jk} \Gamma^{\dagger}$. Recall that since $A$ and $B_{jk}$ depend on the dressed-basis $\ket{k}$ of the spin and bosonic modes, they need to be computed numerically.

\section{Breakdown of Lamb-Dicke regime in dissipative processes}\label{app:diss}
In this appendix we provide additional numerical results to illustrate that, under certain circumstances, some dissipative processes, such as spin dephasing and spontaneous emission and absorption cannot be approximated as $\tilde{D}_{\rm sd}[\rho]\approx \tilde{D}_{\sigma_x}[\rho]$ (Eqs.~(\ref{eq:Dsdapprox}), and~(\ref{eq:Dseapprox}), respectively), as explained in the main text. Indeed, when the Lamb-Dicke condition breaks down, a correct description of dissipation demands the full transformed dissipators as given in Eqs.~(\ref{eq:Dtildesd}) and~(\ref{eq:sigmapm}). In order to illustrate this, we choose $H_{\rm G}$ such that it allows to realize $H_{\rm 1JCM,\eta}$, although the rates are taken to be much larger than the frequencies of $H_{\rm 1JCM,\eta}$ so that the dynamics is essentially governed by $\tilde{D}_{\rm sd}$ or $\tilde{D}_{\rm se}$. Note that we take  $H_{\rm 1JCM,\eta}$, which together with the large dissipation rates, allows us to analyze better how  the approximations performed in the dissipative part spoil the correct functioning of the correspondence between these models.

To inspect the effect of spin dephasing, we consider $\ket{\psi(0)}_{\rm G}=\ket{0}\ket{+}$ as initial state for $H_{\rm G}$ evolving under the presence of spin dephasing and boson losses, with rates $\gamma_{\rm sd}$ and $\gamma_{\rm bl}$. As we show in Figs.~\ref{fig_SpinDephFull}{\it (a)} and {\it (b)}, approximating $\tilde{D}_{\rm sd}[\rho]\approx D_{\sigma_x}[\rho]$ fails to capture the correct equilibrium state for $\eta\gtrsim 0.4$. Indeed, upon taking into account the full  $\tilde{D}_{\rm sd}[\rho]$ (Eq.~(\ref{eq:Dtildesd})), the dynamics is correctly reproduced, as indicated by the low infidelities obtained for the cases plotted in Fig.~\ref{fig_SpinDephFull}. As expected, the crude approximation performed in Eq.~(\ref{eq:Dsdapprox}) breaks down as $\eta$ increases,  and thus resulting fidelities drop significantly (e.g. $F\lesssim 0.8$ for $\eta=0.8$). 

In addition, we also provide results regarding the validity of the Eq.~(\ref{eq:Dseapprox}). For we proceed as before, now choosing $\ket{\psi(0)}_{\rm G}=\ket{0}\ket{e}$ as initial state. As one can observe in  Fig.~\ref{fig_SpinDephFull}{\it (c)} and {\it (d)}, the break down of the Lamb-Dicke condition has a lesser impact in $\tilde{D}_{\rm se}[\rho]$ for intermediate $\eta$ values ($0.4$ or $0.5$) compared to spin dephasing. However, for $\eta=0.8$, a correct functioning of the simulation crucially depends on the inclusion of higher-order terms, such as $a^{n}\sigma^{\pm}$, which are present in Eq.~(\ref{eq:sigmapm}).


%


\end{document}